\documentclass[fleqn, usenatbib]{mnras}

\usepackage{newtxtext,newtxmath}

\usepackage[T1]{fontenc}
\usepackage{ae,aecompl}

\usepackage{graphicx}	
\usepackage{amsmath}	
\usepackage{amssymb}	

\usepackage{caption}
\usepackage{subcaption}
\newcommand*\fig[1]{Figure~\ref{#1}}
\newcommand*\sectn[1]{Section~\ref{#1}}
\newcommand*\eqn[1]{Equation~\ref{#1}}
\newcommand*\ap{a_{\rm p}}
\newcommand*\as{a_{\rm s}}
\newcommand*\bmax{b_{\rm max}}
\newcommand*\sigmax{\sigma_{\rm x}}

\title[Short title, max. 45 characters]{  Planetary Architectures in Interacting Stellar Environments}

\author[Y. Wang et al.]{
Yi-Han Wang,$^{1}$\thanks{E-mail: yihan.wang.1@stonybrook.edu,\newline \indent rosalba.perna@stonybrook.edu}
Rosalba Perna,$^{1,2}$
Nathan  W. C. Leigh$^{3,4}$
\\
$^{1}$Department of Physics and Astronomy, Stony Brook University, Stony Brook, NY, 11794, USA\\
$^{2}$Center for Computational Astrophysics, Flatiron Institute, 162 5th Avenue, New York, NY 10010, USA\\
$^{3}$Departamento de Astronomia, Facultad de Ciencias Fisicas y Matematicas, Universidad de Concepcion, Concepcion, Chile\\
$^{4}$Department of Astrophysics, American Museum of Natural History, Central Park West and 79th Street, New York, NY 10024 
}

\date{Accepted XXX. Received YYY; in original form ZZZ}

\pubyear{2015}

\begin{document}
\label{firstpage}
\pagerange{\pageref{firstpage}--\pageref{lastpage}}
\maketitle

\begin{abstract}
	The discovery of Exoplanetary Systems has challenged some of the theories of planet formation, which assume unperturbed evolution of the host star and its planets. However, in star clusters the interactions with flyby stars and binaries may be relatively common during the lifetime of a planetary system. Here, via high-resolution $N$-body simulations of star-planet systems perturbed by interlopers (stars and binaries), we explore the reconfiguration to the planetary
	system due to the encounters. In particular, via an exploration focused on the strong scattering regime, we derive the fraction of encounters which result in planet ejections, planet transfers and collisions by the interloper star/binary, as a function of the characteristics of the environment (density, velocity dispersion), and for different masses of the flyby star/binary. We find that binary interlopers can significantly increase the cross section of planet ejections and collisions, while they only slightly change the cross section for planet transfers. Therefore, in environments with high binary fractions, floating planets are expected to be relatively common, while in environments with low binary fractions, where the cross sections of planet ejection and transfer are comparable, the rate of planet exchanges between two stars will be comparable to the rate of production of free-floating planets.  

\end{abstract}




\section{Introduction}
The discovery and study of the properties of more than four thousand planets outside of the solar system  
\citep{Wolczan1992,Mayor1995,Bathala2013}
has shown
that our planetary system is not generic, and that some of the ideas developed to explain the characteristics
of the solar planets are not sufficient to account for the larger variety of planetary features. Several observations
have in fact challenged some commonly-made hypotheses in planetary formation. A notable one is the
presence of planets on highly eccentric orbits and  at high relative inclinations \citep{Marcy2003}, both of which
are at odds with the typical assumption of planets forming co-planar and on nearly-circular orbits \citep[e.g.][]{moe18,moe19}.

Another surprise has been the discovery of giant planets with orbits that put them in close proximity with
their host star (the so-called 'hot-Jupiters').  This is again at odds with planet formation theories
which predict that these gaseous giants should form at large separations from their host star, beyond the frost line.
Additionally, some of these giants are found to have misaligned inclination axes, which are again not
generally expected \citep{Winn2010}.  Another challenge to our understanding of planetary formation comes from the
presence of systems with planets on very wide orbits, $\sim 100-10^6$~AU (e.g. \citealt{Lafreniere2011}); the origins of these
are still debated \citep[e.g.][]{moe20}.

The excess number of single-transit planetary systems reported by NASA's {\em Kepler} mission (e.g. \citealt{Lissauer2011,Johansen2012}) is
also believed to be inconsistent with the predictions of modern theories of planetary formation, which predict
that higher multiplicities should be typical (e.g. \citealt{Hansen2013}).
Finally,  recent observations of giant planets around low-mass stars \citep{Morales2019} have further challenged standard
paradigms: according to the core accretion theory,  these types of star-planet systems are difficult to form \citep{Laughlin2004}.

All of the above observations clearly indicate the need for additional mechanisms to operate other than standard planet formation
theories.  There have been many works in the literature in this direction.
Most modifications to the theory have been of an intrinsic nature; that is, internal evolution within the planetary systems.
Dynamical interactions between the planets and the protoplanetary disk in which they form can cause
migration of giants to much smaller orbits (i.e. \citealt{Nelson2000}); this process is further aided by tidal interactions with the
host star \citep[e.g.][]{samsing18}.
Similarly, large eccentricities can be produced via dynamical interactions within the planetary systems
themselves \citep{Ford2003,Ford2006}, albeit the full predicted distribution of eccentricities and semimajor axes for this mechanism have yet
to be theoretically reproduced. Scatterings among planets and/or the Lidov-Kozai mechanism \citep[e.g.][]{hamers18}
can be further held responsible for destabilizing the planetary systems, and eventually lead to a reduction
of the planetary multiplicity \citep[e.g.][]{fregeau06}.

On the other hand, observations of giant planets around small (e.g. \citet{Morales2019}), M-type stars cannot be accounted
for by internal dynamical interactions and secular effects. A possibility is that the giant planet formation occurred as a result of the
onset of the gravitational instability of the young protoplanetary disk, at a time that the disk is still
massive enough relative to the host star \citep{Boss2006}.

While modifications to the standard theory of planet formation based on internal mechanisms alone
are obviously attractive, an increasing number of investigators have started to explore the influence of {\em external}
perturbations on planetary architectures \citep[e.g.][]{hamers19a,hamers19b}.

Dense stellar environments tend to be home to frequent near and even direct
gravitational interactions between pairs of objects in the cluster \citep{hut83b,leonard89,leigh07,leigh11b,leigh13b}.
Such interactions tend to be chaotic in nature, and can lead to the ejection of orbiting bodies,
collisions or the engulfment of orbiting bodies by their host star, alteration of orbital parameters, etc.  \citep[e.g.][]{hut83,bacon96,sills99,leigh12,leigh16}. Hence, one could envisage that, even if there were an
universal birth population of planetary systems, it would become systematically modified over
time by dynamical interactions in dense environments, leading to
the formation of curious orbital architectures unlikely to have formed primordially (e.g.  \citealt{Heggie1996,Laughlin1998,Davies2001,Bonnell2001,Thies2005,fregeau06,Olczak2010,Chatterjee2012,Hao2013,Portegies2015,Li2015,Shara2016,Cai2017,Cai2018,Cai2019,Rice2018,vanelteren2019, Flammini2019}).

External perturbations may have occurred even in field stars like our very own solar system.
In fact, studies of isotopes found in meteorites have shown that, while the solar system was still
forming, it got enriched with many short-lived nuclides (see the review by \citealt{Goswami2000}).  A natural explanation for this
enrichment is provided by the presence in the vicinity of the solar nebula
of either a massive star or of a supernova explosion \citep{Cameron1995}.  Based on these observations,
\citet{Adams2001} suggested that the solar system formed within a stellar group, which they estimated to contain
$N\approx 2000\pm 1100$ members.

Dynamical interactions in dense cluster environments may contribute to alleviating some
of the apparent shortcomings of standard planetary formation, and/or provide alternative
explanations.  \citet{Boley2012}  compared the outcomes of planet-planet scatterings with and without flybys,
and found that the latter influence the distributions of the mutual inclinations in the low/moderate regime.
Additionally, they found that they could induce some giants to migrate inwards into those regions of phase space corresponding to the production of hot Jupiters.
A case for the potentially important role of stellar perturbations in creating hot Jupiters was further made
by \citet{Shara2016}.  Dynamical encounters with flybys have been shown to influence even the orbits
of single planets (e.g. \citealt{Laughlin1998}), while frequent close encounters can be very destructive for planetary systems, even producing free-floating planets
(e.g. \citealt{fregeau06}).   The presence or absence of free-floating planets in cluster environments can be constrained using stellar variability \citep[e.g.][]{gilliland00,weldrake05,weldrake08}. These free floating planets, when transferred to orbit new host stars via dynamical
interactions, could lead to planets on very wide orbits \citep{Perets2012}.

In this paper, via high-resolution $N$-body simulations, we study the statistical
outcomes of flybys from stars and binaries interacting with a simple planetary system made up of a star and a
Jupiter-like planet.  Such systems can either be isolated initially, or part of a binary star system.
For each type of interaction, we compute the cross section for each one of the possible outcomes that
alter the initial planetary architecture.  That is, ejections, collisions and planet transfers, in which the planet is removed from its orbit about its original host star and transferred to orbit an interloper (which could either be a single star or a binary).
We study the dependence on the initial relative velocities, the initial planet separation, and the masses of the stars,
considering both equal masses and unequal ones.  We provide analytical approximations to the cross
sections for the outcomes in which the planet is lost to its birth star either by transfer or by ejection.  Relative to previous studies (in particular \citealt{fregeau06}), we consider both single and binary stars, where the latter provide an additional reservoir of energy and angular momentum relative to the former, via the introduction of additional \textit{orbital} energy and angular momentum.  As we will show in this paper, this can enhance the cross sections for particular outcomes of interest (e.g., planet ejections) relative to the case of an isolated single star interloper.  We further expand the parameter space of initial conditions and extend numerical simulations into a regime not previously accessible using standard regularization techniques alone (e.g., algorithmic regularization), namely extremely low mass ratios (q $\lesssim$ 10$^{-3}$) and high eccentricities (e $\gtrsim$ 0.6) \citep[e.g.][]{fragione18,bonetti20}.

Our study highlights a variety of ways in which dynamical interactions can contribute to creating diversified
planetary architectures, with more possibilities than predicted by the standard scenarios of isolated planetary formation.

\section{Numerical methods}

We perform our scattering experiments with our code
	{\tt SpaceHub} (details in \citealt{Wang2018,Wang2019}), which can handle the extreme mass ratios due to the star-planet systems,
as well as the close encounters, with extremely high precision.
This is achieved via the implementation of cutting-edge chain regularization \citep{Mikkola1993} and on-the-fly round-off error compensation. As a result, the code is able to  treat the extreme cases that with traditional integrators often result in inaccurate results.

\subsection{Scattering configurations}
Depending on the environment, the binary fraction can vary from $\sim1-10\%$ in globular clusters to $\sim50\%$ in the field \citep[e.g.][]{geller08,raghavan10,milone12}; thus, the interactions between single-single stars, single-binary stars and binary-binary stars are dominant in the globular cluster regime over other processes such as triple interactions, whereas interactions involving triples, singles and binaries occur at comparable frequencies in the open cluster and field regimes \citep{leigh13}. In this work, we begin with a general exploration of the parameter space to gain a sense of the interaction rate and cross section for the various outcomes under different conditions. We explore four types of interactions, that is: single star - single star with planet (1+1), binary star - single star with planet (2+1), single star - binary star where one of its components has a planet (1+2) and binary star - binary star with one of its components having a planet (2+2).

\begin{figure}
	\includegraphics[width=0.9\columnwidth]{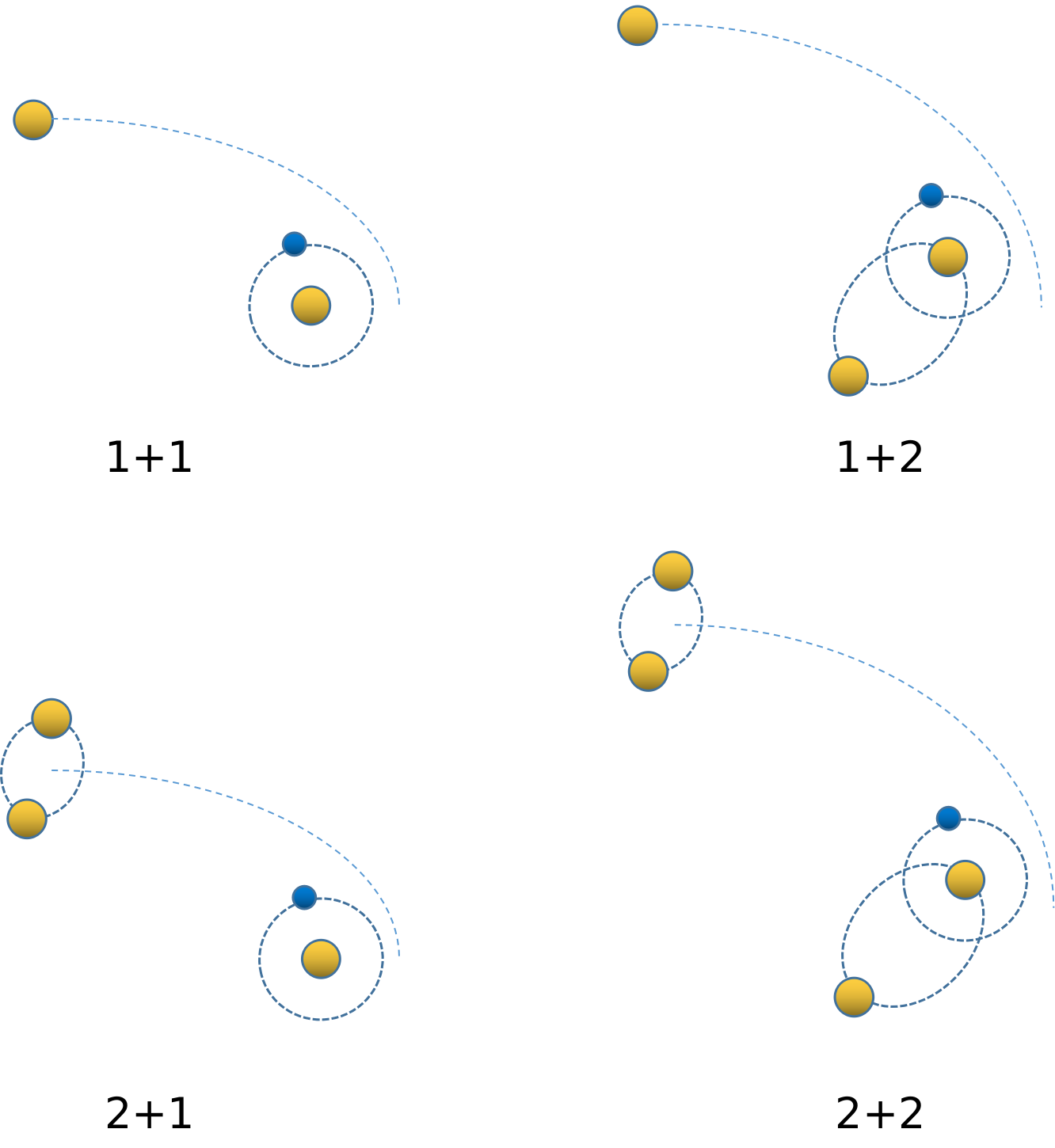}
	\caption{Schematic illustration of the four types of  scattering experiments performed in this paper.}
	\label{fig:schematics}
\end{figure}

\fig{fig:schematics} shows the schematic illustration of the four interactions.

\subsection{General parameter space explorations}
We study the interaction rates and cross sections of 1+1, 1+2, 2+1, 2+2 for planet ejection, transfers and collisions.
For each of our four types of scattering experiments, we compute the interaction rates, and numerically derive
the cross sections, as a function  of the variables
$v_{\infty}$ and $\ap$, where $v_{\infty}$ is the relative velocity at infinity between the scattering objects,
and $a_{\rm p}$ is the orbital separation of the planet from its host star. We explore a range of $v_{\infty}$ values, from 0.1 km~s$^{-1}$ to 30 km~s$^{-1}$, and for $\ap$ from $0.1$ AU to $5$ AU with equally spaced binning. For each pair of $(v_\infty, \ap)$, 1 million scatterings are performed to calculate the interaction rate and cross section for the specific outcomes. The mass of the planet is set to 1 Jupiter mass, and we further vary the mass of the star $M_{\rm s}$ from 1~$M_\odot$ to 0.2~$M_\odot$ to explore the outcome of the dynamical interactions as a function of the mass of the interloper.
We note that, being low-mass stars more numerous than solar type  stars, interactions with these stars are the most frequent ones.
For the separation $\as$ between the stars in the binary, we consider a representative value of
10~AU.

In the 1+2 and 2+2 experiments, the three-body system that includes the planet needs to be stable before the scattering.
To this effect, for this specific configuration we exclude the parameter space in which \citep{Mardling01}
\begin{equation}
	\frac{\as}{\ap} \sim < 3.7\,.
\end{equation}
In total, 400 million scatterings are performed under the automatic task parallel scheme of {\tt SpaceHub}.

For each pair of $(v_\infty, \ap)$, $b^2$ is uniformly drawn from the range $[b, \bmax^2]$, where $b$ is the impact parameter, and $\bmax$ is determined by $v_\infty$ and $\ap$,
\begin{equation}
b_{\rm max} =p_{\rm max}\sqrt{1 + 2\frac{G(M_0+M_1)}{v_\infty^2p_{\rm max}}},
\end{equation}
where $p_{\rm max}$ represents a safe, large enough 'closest approach distance' such that, for all the scatterings with closest approach larger than $p_{\rm max}$, the outcomes are flyby only. 
The phases of the scattered and incident objects are isothermally distributed. For each orbit, $\cos(i)$ is uniformly distributed within the range $[-1,1]$, $\Omega$ is uniformly distributed in $[-\pi,\pi]$, $\omega$ is uniformly distributed in $[-\pi,\pi]$, $M$ is uniformly distributed in $[-\pi, \pi]$, where $i$ is the orbital inclination, $\Omega$ is the longitude of the ascending node, $\omega$ is the argument of periapsis and $M$ is the mean anomaly. The termination time of the integration is set to be one complete flyby (twice the time to pericenter) for positive total energy systems, while we let the system interact several times (we found 10 times to be a safe number for convergence of the results)  for resonant scattering systems with
total negative energy.

\subsection{Classification of the outcomes}
The are several outcomes resulting from the 1+1, 1+2, 2+1 and 2+2 scatterings.  While most of the past literature has
focused on  scatterings of this type between stars alone  (e.g. \citealt{Hills1976,Heggie1975,Mikkola1983,Mcmillan1990,sigurdsson93,bacon96,Giersz2003,Fregeau2003,fregeau06,Samsing2014,Ryu2017a,Ryu2017b}), here our focus is the outcome of a planet orbiting a star as a result
of the interaction.
While the 1+1 case with a planet was studied also by \citet{fregeau06}, here we treat the problem in  greater generality: in particular, we
consider a larger variety of astrophysical scenarios involving also binaries, we compare the numerically-derived cross sections for the various cases and provide
the orbital properties of the planet when transferred for all the cases, as well as the velocity distribution of the ejected planets.

In the post scattering phase, the state of the planet can be any of the following: isolated from all other stars, bound to its original star, and transferred to other stars. Therefore, based on the state of the planet and the configuration of the stars, we classify the outcomes into

\begin{itemize}
	\item \textbf{Ejection:} The planet is ejected from the system, regardless of what is the configuration of the remaining stars.

	\item \textbf{Flyby:} The planet stays bound to its original star, regardless of what is the configuration of the remaining stars.

	\item \textbf{Transfer:} The planet is transferred from its birth star to another star. Due to the initial and end stars' configurations, we divide this case into three subsets.\\
	      \indent \indent 1. \textit{External transfer to single:} The planet is transferred to an isolated single star.\\
	      \indent \indent 2. \textit{External transfer to binary:} The planet is transferred to a binary star and orbits around one component of it.\\
	      \indent \indent 3. \textit{Internal transfer:} The planet is transferred within a binary star, from one component to the other.\\
	\item \textbf{Collision:} Star-star collision and star-planet collision.

	\item \textbf{Others:}  Other configurations (such as the ones involving the formation of triples) that we are not interested in. These constitute less than 1\% of the total outcomes.
\end{itemize}

Here we  derive the interaction rate and cross section of the outcomes in which the planetary architecture is reconfigured by the scattering via any of the processes of ejection, transfer and collision. We study the properties of the planet in the new configuration, and discuss it within the context of various astrophysical environments.
Our results are meant to help further our understanding of the observed wide phenomenology of planetary
architectures.

\subsection{Outcome ratios and cross sections}
For a given interaction
velocity at infinity, we perform scattering experiments with $b^2$ uniformly distributed in the range $[0, \bmax^2]$.
Therefore, if we run $N_{\rm tot}$ scattering experiments and obtain $N_{\rm x}$
outcomes, we can estimate the outcome ratio to be
\begin{equation}
	f_{\rm x} = \frac{N_{\rm x}}{N_{\rm tot}}\,.
\end{equation}

The corresponding cross section $\sigmax$ for the given outcome  can then be calculated from (see e.g. \citealt{fregeau04})

\begin{equation}
	\bar{\sigmax} = \pi \bmax^2\frac{N_{\rm x}}{N_{\rm tot}},
\end{equation}
with standard deviation
\begin{equation}
	\Delta \sigmax = \pi \bmax^2\frac{\sqrt{N_{\rm x}}}{N_{\rm tot}}\,.
\end{equation}

 This calculation is independent of the choice of $\bmax$ only if $\bmax \ge b_{\rm c}$, where $b_{\rm c}$ is the critical impact parameter such that, for $b>b_{\rm c}$, all the outcomes are flyby only. Therefore, for scatterings with $b_{\rm c}<b<\bmax$, $N_{\rm x}=0$. Then, we have,
\begin{eqnarray}
	\bar{\sigmax} &=& \pi \bmax^2\frac{N_{\rm x}}{N_{\rm tot}}\nonumber \\
	&=&\pi b_{\rm c}^2\frac{N_{\rm x}(b<b_{\rm c})}{N_{\rm tot}(b<b_{\rm c})} + \pi (\bmax^2 - b_{\rm c}^2)\frac{N_{\rm x}(b_{\rm c}\le b \le \bmax)}{N_{\rm tot}(b_{\rm c}\le b \le \bmax)}
	\nonumber \\
	&=&\pi b_{\rm c}^2\frac{N_{\rm x}(b<b_{\rm c})}{N_{\rm tot}(b<b_{\rm c})}\,.
\end{eqnarray}
Note that the second step of the above equation assumes that the sampling density is constant for any $b<b_{\rm max}$.

\subsection{Cross section calculation with velocity dispersion}
Note that the numerical method for the computation of the
cross section  is a function of $\bmax$, which is in turn  a function of $v_{\infty}$. The cross section calculation needs $\bmax$ to be fixed. The usual way this is done in the literature is by fixing  $v_{\infty}$. However, the velocity dispersion, which scales linearly with $v_{\infty}$ in astrophysical environments,  usually obeys a distribution. For simplicity, a Maxwellian distribution with 1D dispersion is usually adopted,
\begin{equation}
	f(v_\infty) \propto v_{\infty}^2 e^{\frac{v_\infty^2}{2\sigma^2}}\,.
\end{equation}

Thus, the cross section for a given velocity distribution is
\begin{eqnarray}\label{eq:cross-v-dist}
	\langle \sigmax\rangle &=&\int\sigmax(v_\infty)f(v_\infty){\rm d}v_\infty\nonumber \\
	&=&\int\pi \bmax^2(v_\infty)\frac{N_{\rm x}(v_\infty)}{N_{\rm tot}(v_\infty)}f(v_\infty){\rm d}v_\infty \nonumber \\
	&=&\int\pi \bmax^2(v_\infty)\frac{N_{\rm x}(v_\infty)}{N_{\rm tot}}\frac{N_{\rm tot}f(v_\infty)}{N_{\rm tot}(v_\infty)}{\rm d}v_\infty \nonumber\\
	&=&\frac{1}{N_{\rm tot}}\int\pi \bmax^2(v_\infty)N_{\rm x}(v_\infty)\frac{N_{\rm tot}(v_\infty)}{N_{\rm tot}(v_\infty)}{\rm d}v_\infty \nonumber\\
	&=&\frac{\pi}{N_{\rm tot}}\int \bmax^2(v_\infty)N_{\rm x}(v_\infty){\rm d}v_\infty\,,
\end{eqnarray}
where $N_{\rm x}(v_\infty)$ and $N_{\rm tot}(v_\infty)$ are, respectively, the number of outcome $x$ and the total number of scatterings for a specific $v_\infty$. With $v_\infty$  drawn from the given distribution, then the impact parameter $b$ is chosen to be within the range $[0, \bmax(v_\infty)]$ with $b^2$ uniformly distributed. We perform the scattering experiments $N_{\rm tot}$ times. Then, the cross section can be calculated via \eqn{eq:cross-v-dist} using discrete integrations.

\section{Results}

\subsection{Critical velocities}\label{sec:vc}
The critical velocity $v_{\rm c, E}$, defined as the value for which the total energy of the system is zero, marks the border between hard/soft scatterings \citep{Heggie1975}. If $v_\infty < v_{\rm c,E}$, the binaries will on average harden after the scattering, while if $v_\infty > v_{\rm c,E}$, the binaries will on average soften. This is accurate for equal mass scattering, where the masses of the scattering objects are almost equal. However, \citet{Hills1989} and \citet{Hills1990} (see also \citealt{fregeau06}) suggested that for extreme unequal mass scatterings, this   boundary is more accurately provided by $v_{\rm c,I}$, considering the relative velocity between the scattered object and the intruder. They suggested the use of the terminology 'fast/slow boundary'. If $v_\infty < v_{\rm c, I}$, the relative velocity between scattered object and intruder can be zero, and hence long interacting times can be achieved. If $v_\infty > v_{\rm c, I}$, the relative velocity between the scattered object and the intruder will never reach zero, and thus the interacting time will be short.

In 3-body scatterings (1+1 in this paper), if the scattered object is the planet, the upper limit of the velocity of the scattered object in its host star's reference frame is 
\begin{equation}
v_{\rm p}=\sqrt{ \frac{G(M_{*1}+M_{\rm p})}{a_{\rm p}}},
\end{equation}
and the lower limit of the intruder speed (which is a single star) is $v_\infty$. Therefore, the critical velocity $v_{\rm c, I}$ that makes it impossible for a zero relative velocity between the scattered object and intruder is the velocity where the two limits join. This is exactly the orbital velocity of the planet $v_{\rm p}$. Thus, in the literature, this critical velocity $v_{\rm c, I}$ is also identified with $v_{\rm orb}$.
However, for more complicated scatterings (1+2, 2+1 and 2+2 in this paper), we define this 'interacting time'-based critical velocity more precisely. 

For 1+2 scatterings, the upper limit of the velocity of the planet is $v_{\rm p} + \frac{1}{2}v_{\rm orb,h}$, where $v_{\rm orb,h}$ is the orbital velocity of the host binary. The lower limit of the intruder is $v_{\infty}$. Therefore,
\begin{equation}
v_{\rm c, I, 1+2} = v_{\rm orb,p}+\frac{1}{2} v_{\rm orb,h}\,.
\end{equation}
For 2+1 scatterings, the upper limit of the velocity of the planet is $v_{\rm p}$, and the lower limit of the intruder is $v_{\infty} - v_{\rm orb, i}$. Therefore,
\begin{equation}
v_{\rm c, I, 2+1} = v_{\rm orb,p}+\frac{1}{2} v_{\rm orb,i}\,.
\end{equation}
Last, for 2+2 scatterings, the upper limit of the velocity of the planet is $v_{\rm p} + \frac{1}{2}v_{\rm orb,h}$, and the lower limit of the intruder is $v_{\infty} - v_{\rm orb, i}$. Therefore,
\begin{equation}
v_{\rm c, I, 2+2} = v_{\rm orb,p}+\frac{1}{2} v_{\rm orb,h} +\frac{1}{2} v_{\rm orb,i}\,.
\end{equation}

In this paper for $a_{\rm p}=1$ AU, $v_{\rm p}\sim30$~km~s$^{-1}$, $v_{\rm orb,h}=v_{\rm orb,i}\sim 13.4$~km~s$^{-1}$. The corresponding values of $v_{\rm c,I}$  (30~km~s$^{-1}$ for 1+1,   36.7~km~s$^{-1}$ for 1+2 and 2+1, 43.4~km~s$^{-1}$ for 2+2) are much larger than our exploration region in the parameter space where $v_\infty\sim$ [0.1, 30]~km~s$^{-1}$.

\subsection{Cross section and outcome rate as a function of $\ap$ and $v_\infty$}


\begin{figure}
	\includegraphics[width=\columnwidth]{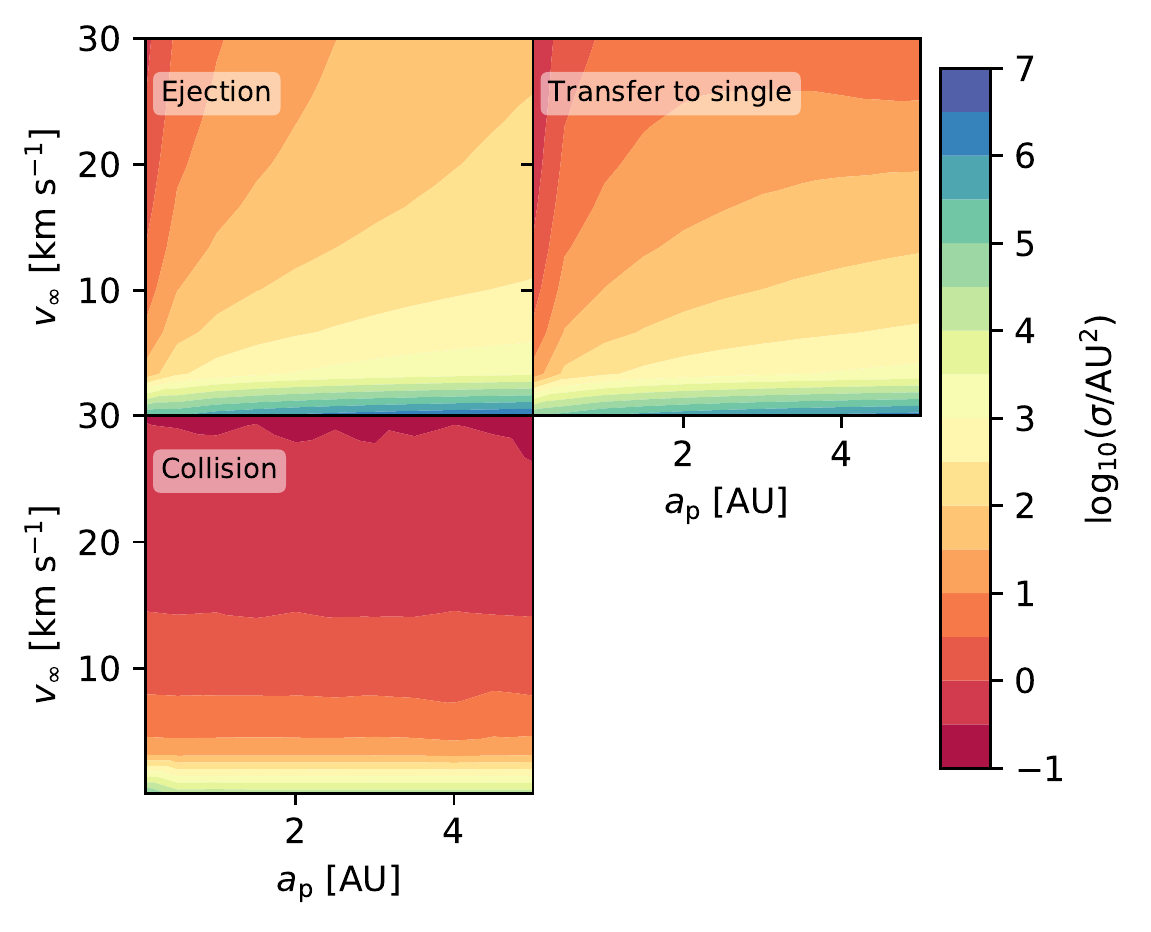}
	\caption{cross sections for planet ejections, planet transfers and collisions for 1+1 scatterings, as a function of $v_\infty$ and $\ap$. The mass of star is $M_{*1}=M_{*2}=1M_\odot$.}
	\label{fig:1+1}
\end{figure}

\begin{figure}
	\includegraphics[width=\columnwidth]{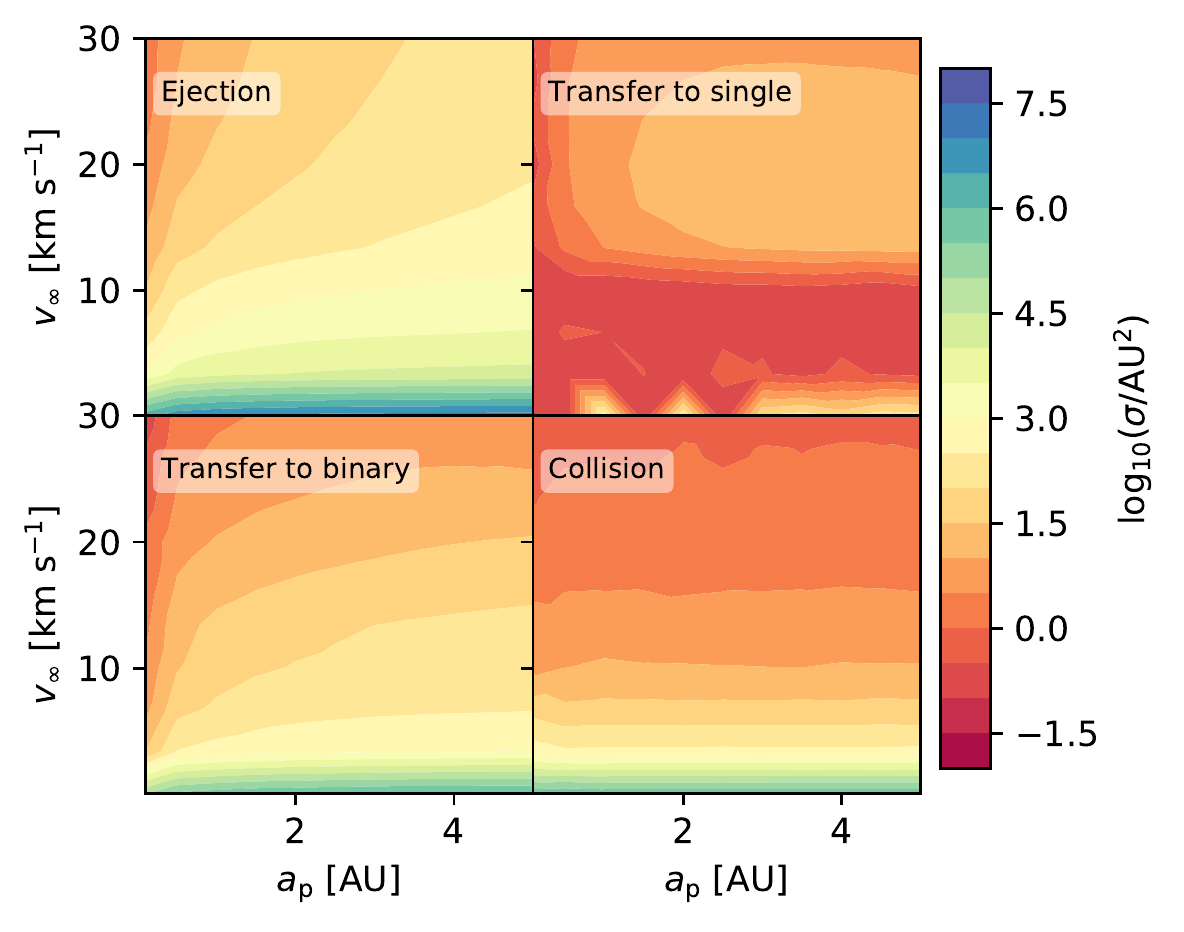}
	\caption{cross sections for planet ejections, planet transfers, and collisions for 2+1 scatterings as a function of $v_\infty$ and $\ap$. The mass of star is $M_{*1}=M_{*2}=1M_\odot$.}
	\label{fig:2+1}
\end{figure}

\begin{figure}
	\includegraphics[width=\columnwidth]{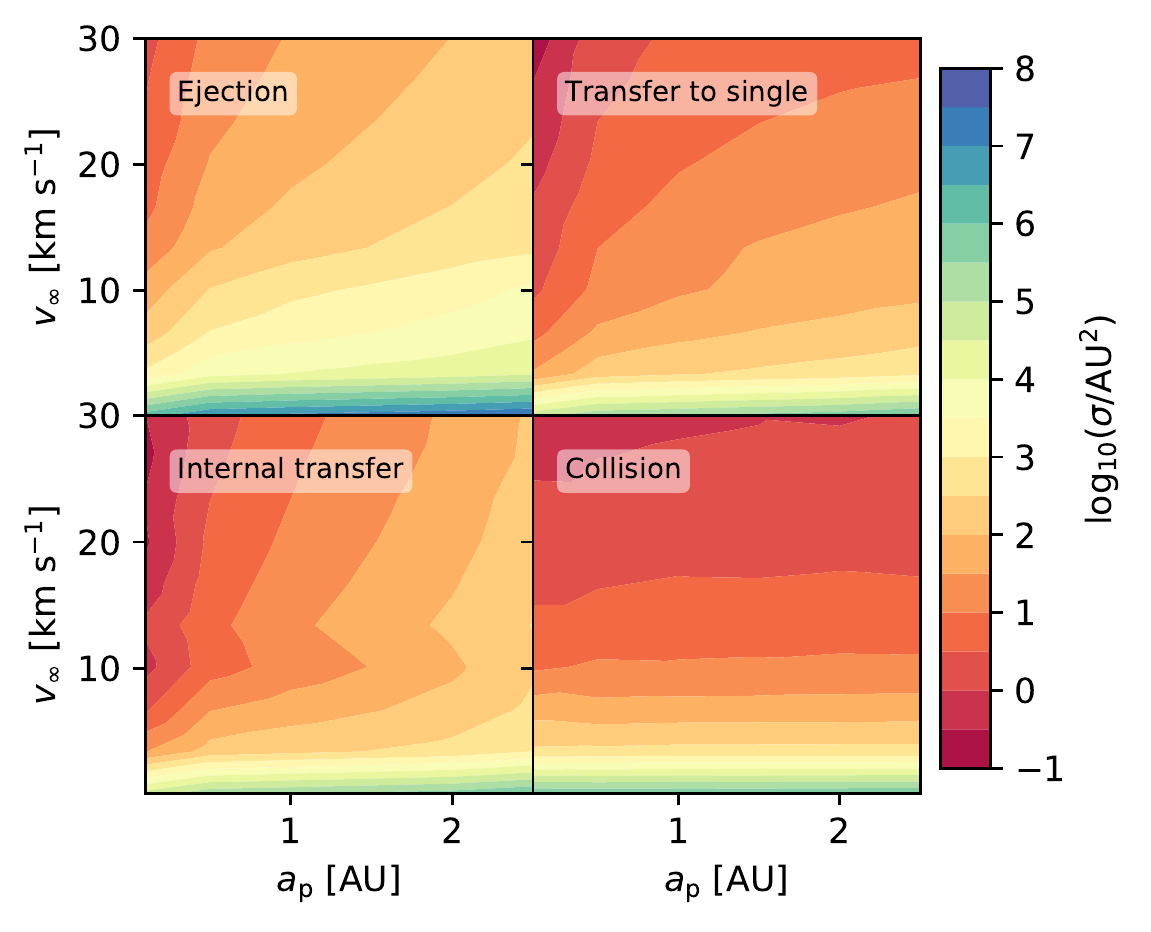}
	\caption{cross section for planet ejection, planet transfer and collision for 1+2 scattering as a function of $v_\infty$ and $\ap$. The mass of star is $M_{*1}=M_{*2}=1M_\odot$.}
	\label{fig:1+2}
\end{figure}

\begin{figure}
	\includegraphics[width=\columnwidth]{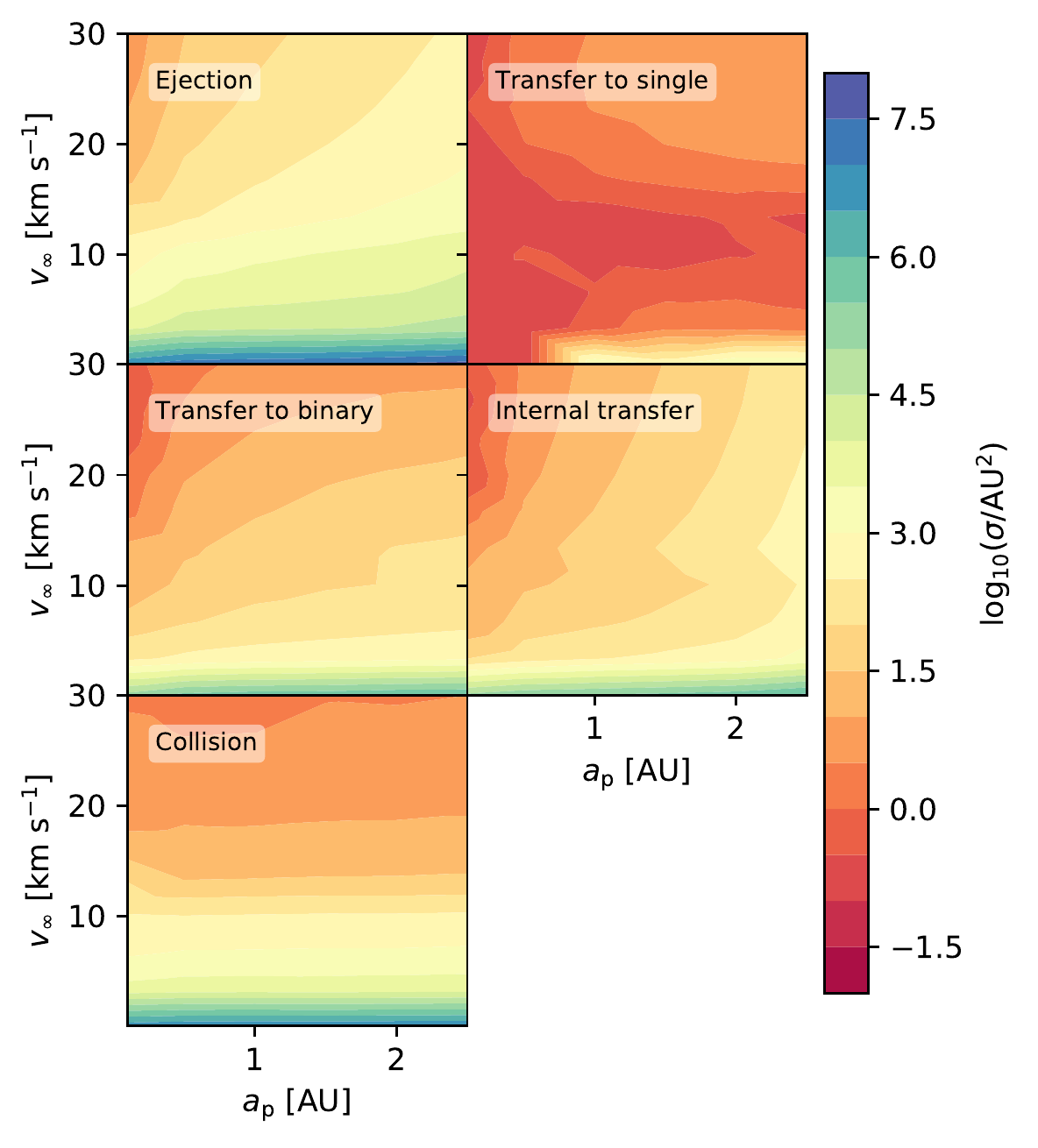}
	\caption{cross section for planet ejection, planet transfer, planet transfer and collisions for 2+2 scatterings as a function of $v_\infty$ and $\ap$. The mass of star is $M_{*1}=M_{*2}=1M_\odot$.}
	\label{fig:2+2}
\end{figure}


\fig{fig:1+1}, \fig{fig:2+1}, \fig{fig:1+2} and \fig{fig:2+2} show the cross sections for  planet ejections, planet transfers and collisions for 1+1, 1+2, 2+1 and 2+2 scatterings with $M_{\rm s} = 1M_\odot$ and $\as=1$ AU. It can be seen that all the cross sections become independent of $\ap$ as $v_\infty$ drops under 3 km~s$^{-1}$. This indicates that in the open cluster regime and for those globular clusters with small velocity dispersions, the interaction outcome is independent of the planet separation if the planet is not a hot Jupiter ($\ap<0.1$ AU). From the comparison between 1+1 and 2+1 scatterings, it is not surprising that the cross sections for ejections and collisions are significantly increased by an order of magnitude in the 2+1 scatterings.  This is due to the larger geometrical cross section of a binary star relative to singles of comparable mass.

\begin{figure}
	\includegraphics[width=\columnwidth]{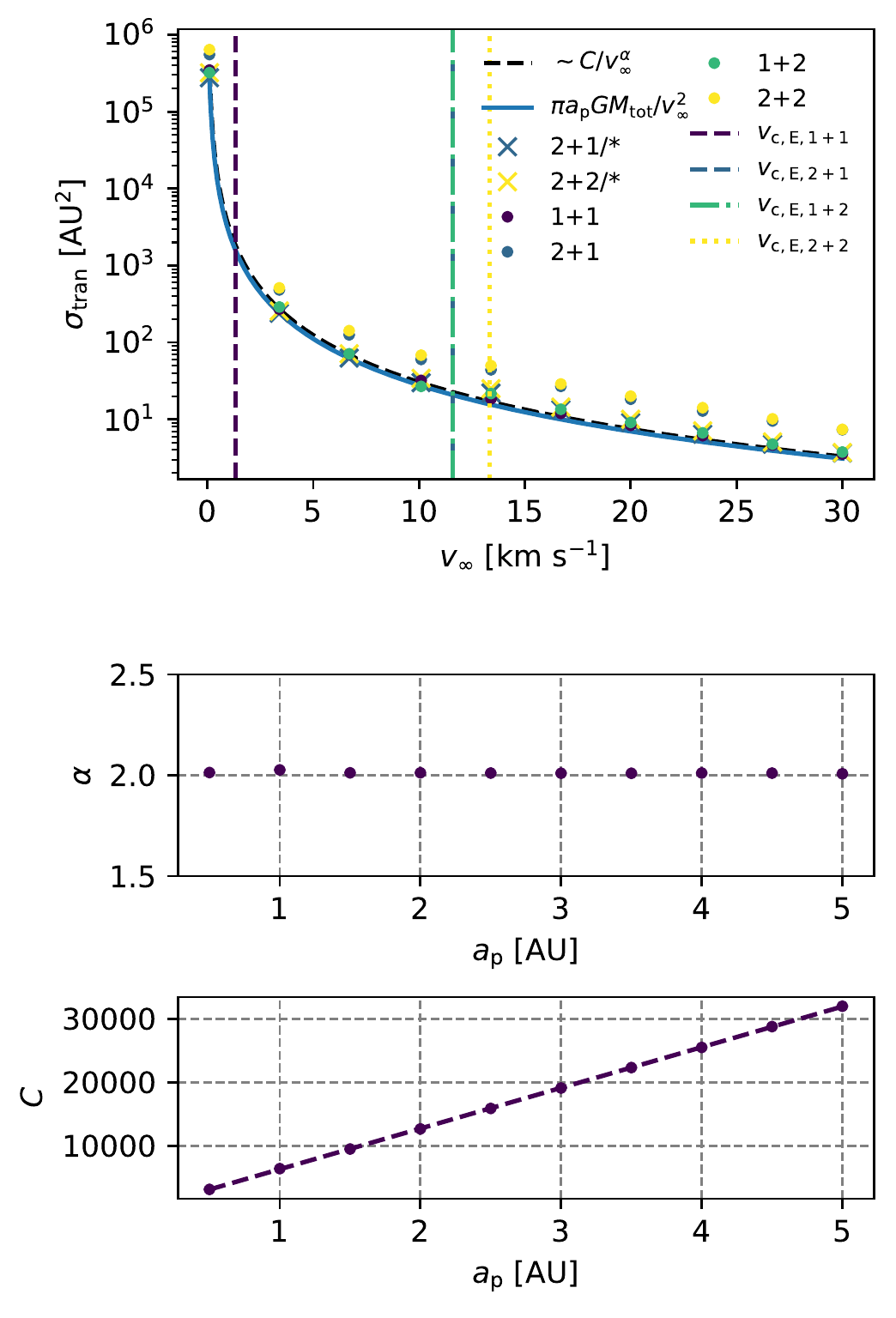}
	\caption{The upper panel shows the cross section for planet transfers as a function of $v_\infty$ with $\ap=2.5$ AU for different types of scatterings. The bottom panel shows the best fit parameters for the 1+1 case. The vertical  lines show the critical velocities for different types of scatterings. }
	\label{fig:transfer}
\end{figure}

An interesting outcome of our simulations  is the fact
that the cross section for transfers (the sum of external transfers to single, external transfers to binary and internal transfers) remains almost the same for the four different configurations and only slightly depends on  $\ap$. \fig{fig:transfer} displays this feature for 1+1, 1+2, 2+1 and 2+2 with $\ap=1$~AU.
We fit the cross section with a power law
\begin{equation}
	\sigma_{\rm tran} = C/v_\infty^\alpha = k\ap/v_\infty^\alpha\,,
\end{equation}
where $k$ and $\alpha$ are the two fitting parameters.  It is indicatively shown that the cross section $\sigma_{\rm tran}$ for both the 2+1 and 2+2 case is  slightly higher than for 1+1 and 1+2. However, we find that, if we normalize  $\sigma_{\rm tran}$ by the number of objects in the capturer (i.e. 2 in 2+1 and 2+2, and 1 in 1+1 and 1+2), the normalized cross sections are pretty much identical. Thus, for planet transfers, we can treat the binary intruder as two independent  single stars. This result also indicates that the binary host star does not affect much the process of planet transfer.

The vertical lines show the 'energy-based' critical velocities $v_{\rm c,E}$ for different scatterings. We see that those  velocity values do not accurately indicate the boundary of the interaction regime for planet-oriented outcomes in such a high mass ratio scattering process as discussed in \sectn{sec:vc}. However, we do note a tiny bump around $v_{\rm c, E, 2+2}$. In fact, in the term $v_{\rm c, E, 2+2}$, the binding energy of the binary star is dominant over the binding energy of the planet. Therefore, this critical velocity is more reflective of the boundary of the interaction regime for the star-star interaction, independently of the presence of the planet. We have found that binary stars play an insignificant role in helping the planet exchange process, and hence we can practically treat this mechanism as the result of interaction with two isolated stars. We believe that this bump indicates the extent to which the planet transfer depends on the configuration of the binary stars in the outcome, but this dependence is almost negligible here.

The lower panel of \fig{fig:transfer} shows the fitting results, where $k$ is found to be $3306$ and $\alpha$ has a best value of $2.03$. Thus, the cross section for transfers obeys the theoretically expected -2 power law
\begin{equation}
	\sigma_{\rm tran} = 3306 \bigg(\frac{M_{\rm s}}{M_\odot}\bigg)\bigg(\frac{\ap}{\rm AU}\bigg)\bigg(\frac{v_\infty}{\rm km~s^{-1}}\bigg)^{-2}{\rm AU}^2 \,,
\end{equation}
regardless of the type or configuration of the scattering (i.e., 1+1, 1+2, 2+1, 2+2).
It is not surprising that we obtain a power law $\approx -2$, since
\begin{eqnarray}
	\sigma = \pi  \bmax^2\frac{N_{\rm cap}}{N_{\rm tot}}\propto p^2 + 2\frac{GM_{\rm tot}p}{v_\infty^2}\,,
\end{eqnarray}
where $p$ is the distance at closest approach. At the low $v_\infty$ limit, $\sigma\sim 1/v_\infty^2$. This $p^2$ term in the equation also explains the deviation in the high-end tail of \fig{fig:transfer}, as it becomes comparable to the $1/v_\infty^2$ term.

We believe that in the range $v_\infty\in[0.1,30] $ km~s$^{-1}$, which covers most of the relevant parameter space for realistic astrophysical environments, our fits provide a fair approximation to the cross section of transfers found from our numerical experiments.


\begin{figure}
	\includegraphics[width=\columnwidth]{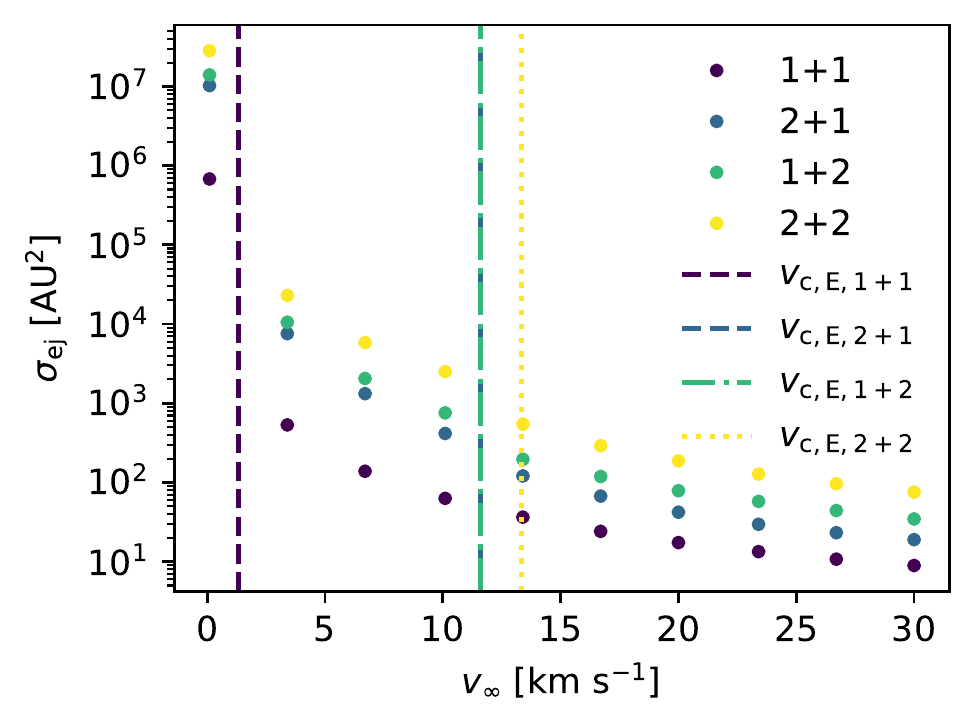}
	\includegraphics[width=\columnwidth]{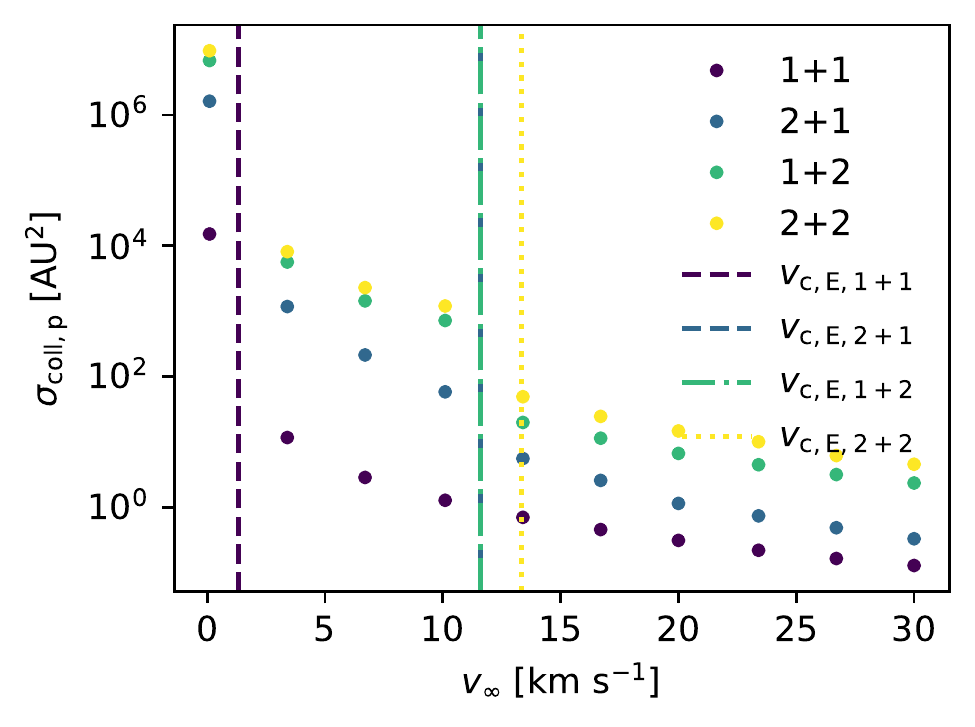}
	\includegraphics[width=\columnwidth]{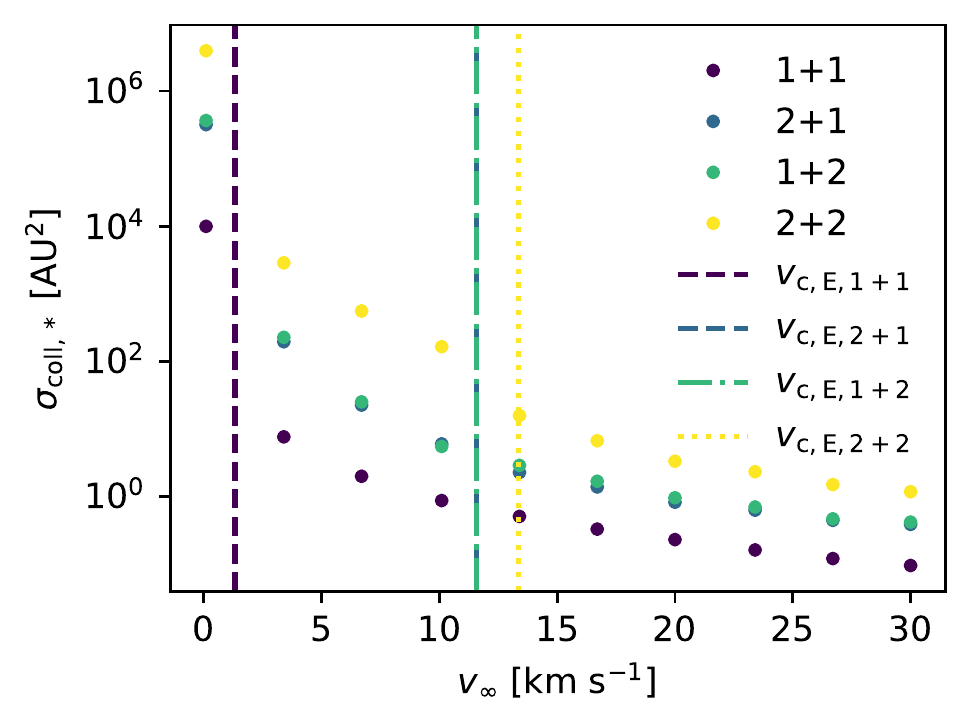}
	\caption{Cross section of planet ejections and collisions (including star-star and star-planet collisions) as a function of $v_\infty$ for different types of scatterings.}
	\label{fig:sigma-others}
\end{figure}

\fig{fig:sigma-others} shows the cross section of planet ejection and collisions (star-star collision and star-planet collision) for 1+1, 1+2, 2+1 and 2+2. The presence of the binary in 1+2, 2+1 and 2+2 scatterings, which increases the geometrical cross section, result in a significantly larger  cross section for ejections and collisions.

One should notice that the geometrical cross sections for 1+2 and 2+1 are basically the same, but the cross section for planet ejection and collision in the 1+2 case, where the planet is orbiting around one component in a binary star, is higher than the 2+1 case. Indeed, what matters here is the geometrical cross section of the planet. In the 1+2 case, the planet orbits around one component of the binary star, while its host star orbits around the center of mass of the binary.  The effective area of the surface consisting of all the possible positions of the planet in the 1+2 case is much larger than the area of the surface in the 2+1 case. In fact, the geometrical cross section of the planet in the 1+2 case is $\sim \pi (\as+\ap)^2$, while the geometrical cross section of the planet in the 2+1 case is only $\sim \pi \ap^2$.

 In Fig.\ref{fig:sigma-others}, we indicate the 'energy-based' critical velocities ($v_{\rm c,E,2+1}$, $v_{\rm c,E,1+2}$ and $v_{\rm c,E,2+2}$) which generally reflect the effect of different configurations in the comparable mass ratio scattering.  
 Here, however, for the 
 planet ejection case of the 2+2 interactions, the star-planet collision of the 2+1, 1+2 and 2+2 cases, and  the star-star collision of the 2+2 encounter, the outcome is entangled with the configurations of the stars, and hence we see a slight dependence of the cross sections on those critical velocities, and
 in particular we note the inflection point corresponding to $v_{\rm c,E,2+2}$.
 Thus, these 'energy based' critical velocities which classify the scattering configurations in the comparable mass case, also classify the regimes of these process here with a planet.

In the regime $v_\infty < v_{\rm c, E}$, if we treat the planet as a test particle, we are in the hard binary regime, where the binary star tends to harden after the scattering. On the other hand, for $v_\infty > v_{\rm c, E}$, the binary star tends to soften. In the hard binary regime, the potential energy release gives a higher chance for the planet to be ejected, thus resulting in a larger planet ejection cross section in this regime.

\subsection{Post-scattered orbital properties of the planet}
In this section we investigate the properties of those objects left over post scattering. These parameters include $v_\infty^\prime$, $v_{\rm ej}$, $\ap^\prime$ and $e_{\rm p}^{\prime}$. They are relevant to planet transfers and ejections. Here, $v_\infty^\prime$ is the final relative velocity between the centre of mass of the left-over star-planet system and the newly formed single star;   $v_{\rm ej}$ is the velocity of the ejected planet at infinity;  $\ap^\prime$ and $e_{\rm p}^\prime$ are, respectively, the semi-major axis and eccentricity of the newly formed orbital parameters of the planet,  if it remains bound.

It is obvious that the internal transfer is quite different from the external transfer to the single and the external transfer to the binary. In the external transfer cases (both to the single and the binary), after  ionization, the planet is recaptured by the hyperbolic incident object. However, in the internal transfer case, the planet, after being ionized, is recaptured by the other star component of the binary, which is in a circular orbit. The difference in speed between the hyperbolic incident object (in the external scenario) and the circular orbital component (in the internal scenario) affects the orbital properties of the transferred planet. Therefore, we display the properties of the external transfer to the single/binary and internal transfer independently.

\begin{figure*}
	\includegraphics[width=2\columnwidth]{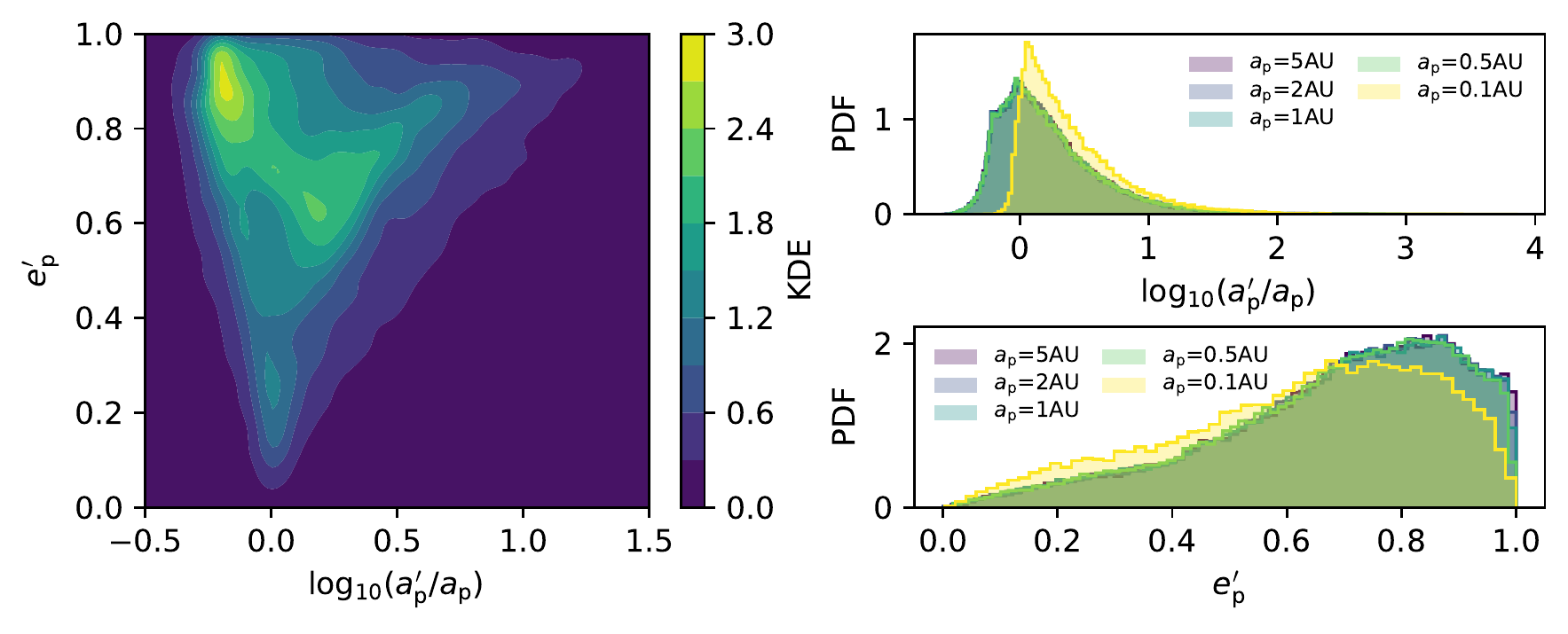}
	\caption{Orbital properties of the external transferred planet in 1+1 scatterings for $v_\infty=0.1$~km~s$^{-1}$. \textit{Left panel:} each subplot shows the KDE of the $\ap^\prime$ and $e_{\rm p}^\prime$ values for initial $\ap=1$~AU. \textit{Right panels:} One-dimensional PDF of the $\ap^\prime$ and $e_{\rm p}^\prime$ values shown separately with different $\ap$.}
	\label{fig:post-properties}
\end{figure*}

\subsubsection{Semi-major axis and eccentricity of external transfer planets}
For planet external transfer in cases other than 1+1, the required interacting distance between the donating star and the accepting star for planet transfer is of the same order of magnitude of the original planet separation $\ap$,  which is usually rather smaller than the binary star separation. Thus, only one component of the accepting binary and the donating binary contribute to the planet relocation. This is further verified by the similar cross section in \fig{fig:transfer}. The orbital properties of the transferred planet for 1+1, 2+1, 1+2 and 2+2 also show similar distributions (see Appendix). Therefore, we only display the distributions of the 1+1 case here to remain concise in the main paper text.

In fact, the new semi-major axis and eccentricity can be estimated from the Keplerian relations
\begin{eqnarray}
	\ap^\prime &\sim& -\bigg( \frac{{\rm d}v^2}{G(M_{*2}+M_{\rm p})}-\frac{2}{{\rm d}r}\bigg)^{-1}\\
	\mathbf{e}_{\rm p}^\prime &\sim&
	(\frac{{\rm d}v^2}{G(M_{*2}+M_{\rm p})}-\frac{1}{{\rm d}r})\mathbf{{\rm d}r} - \frac{{\rm d}\mathbf{v} \cdot {\rm d}\mathbf{r}}{G(M_{*2}+M_{\rm p})} {\rm d}\mathbf{v}\,,
\end{eqnarray}
where ${\rm d}\mathbf{r}$ and ${\rm d}\mathbf{v}$ are, respectively,  the relative position and velocity between the incident star and the planet at the moment of
closest approach.
Since $\mathbf{e}_{\rm p}^\prime$ is a constant of motion, a simplification  of this equation can be made by taking ${\rm d}\mathbf{r}$ and ${\rm d}\mathbf{v}$ to be perpendicular. We obtain
\begin{equation}
	e_{\rm p}^\prime \sim \frac{{\rm d}r{\rm d}v^2}{G(M_{*2}+M_{\rm p})}-1\,.
\end{equation}
In order to make a planet transfer rather than an ejection, $e_{\rm p}^\prime$ must be smaller than $1$. Therefore ${\rm d}r$ and ${\rm d}v$ must be constrained by
\begin{equation}\label{eq:condition}
	{\rm d}r{\rm d}v^2<2G(M_{*2}+M_{\rm p})\,.
\end{equation}
The relative position ${\rm d}r$ is less sensitive to the constraint while the relative velocity ${\rm d}v$ will significantly affect the new eccentricity. The relative velocity ${\rm d}r$ in this scattering scenario is
\begin{equation}
	{\rm d}\mathbf{r}=\mathbf{v}_{\rm re}-\mathbf{v}_{\rm p}\,,
\end{equation}
where $\mathbf{v}_{\rm re}$ is the velocity (relative to the original star) of the incident star at the closest approach, and $\mathbf{v}_{\rm p}$ is the orbital velocity of the planet. For strictly prograde   and retrograde orbits,
\begin{equation}
	{\rm d}v=|\mathbf{v}_{\rm re}|\pm|\mathbf{v}_{\rm p}|\,.
\end{equation}
Therefore, for prograde transfers, where ${\rm d}v=||\mathbf{v}_{\rm re}|-|\mathbf{v}_{\rm p}||$, the mean value of the new $e_{\rm p}^\prime$ is expected to be smaller than for retrograde transfers, where ${\rm d}v=||\mathbf{v}_{\rm re}|+|\mathbf{v}_{\rm p}||$.

The three subplots in \fig{fig:post-properties} with $v_\infty = 0.1$~km~s$^{-1}$, $3.4$~km~s$^{-1}$ and $10.1$~km~s$^{-1}$, which, respectively,  cover the velocity dispersion of the open cluster, globular cluster and field stars, are pretty similar. This indicates that the velocity dispersion of the environment in this range for solar mass stars is insensitive to the post-scattered orbital properties of the transferred planet.

The left panel of each subplot in \fig{fig:post-properties} clearly shows the two peaks  corresponding to
the two different modes of planet swap in the
($\ap^\prime$- $e_{\rm p}^\prime$) kernel density,  for $M_{*1}=1M_\odot$, $M_{*2}=1M_\odot$,
$\ap=1$~AU and $v_\infty=0.1$~km s$^{-1}$ . The left peak shows the retrograde transfer while the right peak shows the prograde one.
The same reasoning can also be applied by noting that $\ap^\prime>0$ in order to capture the planet in a closed orbit ($\ap^\prime<=0$ is a parabolic/hyperbolic orbit), and hence conclude that the prograde transfer results in larger semi-major axes while the retrograde transfer results in smaller semi-major axes, which is clearly shown in the left panel of \fig{fig:post-properties}.

The right panels of each subplot in \fig{fig:post-properties} show the PDF of the collapsed distribution of $\ap^\prime$ and $e_{\rm p}^\prime$ for different initial values of $\ap=5,2,1,0.5$~AU. The two panels indicate that, while the distributions of  $\ap^\prime$ and $e_{\rm p}^\prime$ (which result from  superposition of two peaks), are almost identical, however, when $\ap=0.1$AU (tightly bound planet), the left peak which corresponds to the retrograde transfer becomes extremely faint (essentially missing). Indeed, for swaps in retrograde orbits,
\begin{equation}
	{\rm d}v=||\mathbf{v}_{\rm re}|+|\mathbf{v}_{\rm p}||\sim ||\mathbf{v}_{\rm re}|+(GM_{*1}/\ap)^{1/2}|\,.
\end{equation}
A small $\ap$  increases this quantity and makes the condition in Eq.(\ref{eq:condition}) hard to satisfy for a given distribution of ${\rm d}r$ that is determined by $v_\infty$ and the masses. Thus, planet transfers from this channel are largely suppressed.

\subsubsection{Semi-major axis and eccentricity of internal transfer planets}
There is a non-negligible fraction of internal planet transfers in the 1+2 and 2+2 cases, although it is small compared to that of external transfers. Similarly to the case of the external transfer to the  single/binary, we find that the post-scattered planet orbit shows a very  weak dependence on the environment velocity dispersion and on the  interaction type (1+2 or 2+2) (see Appendix).

\begin{figure*}
	\includegraphics[width=2\columnwidth]{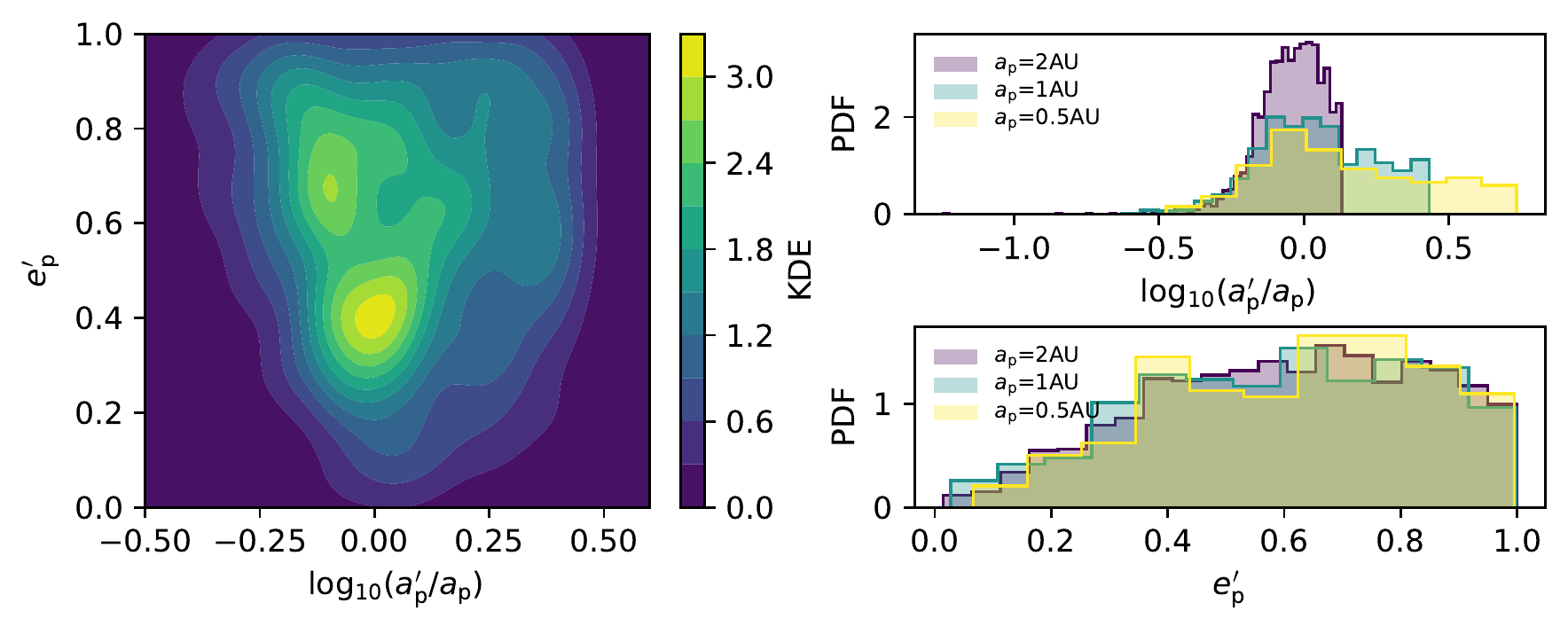}
	\caption{Orbital properties of the internal transfer planet in 2+1 case for $v_\infty=30.0$~km~s$^{-1}$. The KDE of the $\ap^\prime$ and $e_{\rm p}^\prime$ for initial $\ap=1$~AU. \textit{Right panels:} The one-dimensional PDF of the $\ap^\prime$ and $e_{\rm p}^\prime$ separately with different $\ap$.}
	\label{fig:trans-post-properties}
\end{figure*}

\fig{fig:trans-post-properties} shows the orbital properties of the transferred planet for $v_\infty = 30$~km~s$^{-1}$. The left panel shows the 2D KDE of the $\ap^\prime$ and $e_{\rm p}^\prime$, while the right panels show the collapsed 1D PDE. The distribution in the upper right panel is truncated by the three body stability criteria. In a binary star with the planet orbiting one component of it, the semi-major axis of the planet orbit needs to be small enough compared to the semi-major axis of the binary star, to make the triple stable. Therefore, for internal transfer planets, there is a hard cut upper limit to the semi-major axis $\ap^\prime$. This upper limit is higher for wider binary stars as indicated in the plot.

\subsubsection{Velocity of transferred planet-star system}
The scattering experiments are terminated at the point at which the new star-planet system can be regarded as an isolated binary (tidal factor $\delta \sim 10^{-5}$). Thus, we take the semi-major axis $\ap^\prime$ and the eccentricity  $e_{\rm p}^\prime$ as the new orbital parameters 
 at  infinity. The new relative velocity $v_\infty^\prime$ is calculated from

\begin{eqnarray}
	v_\infty^\prime=\frac{M_{*1}+M_{*2}+M_{\rm p}}{M_{*,1}}\bigg(v_{\rm cm}^2 -\frac{2GM_{*1}}{r}\frac{M_{*1}}{M_{*1}+M_{*2}+M_{\rm p}} \bigg)^{1/2}\,,
\end{eqnarray}
where $v_{\rm cm}$ is the centre of mass velocity of the new star-planet system in the centre of mass reference frame and $r$ is the relative distance between the new system and the planet-donating star. The variables $v_{\rm cm}$ and $r$ are measured at the time of the simulation termination.
\begin{figure}
	\includegraphics[width=1\columnwidth]{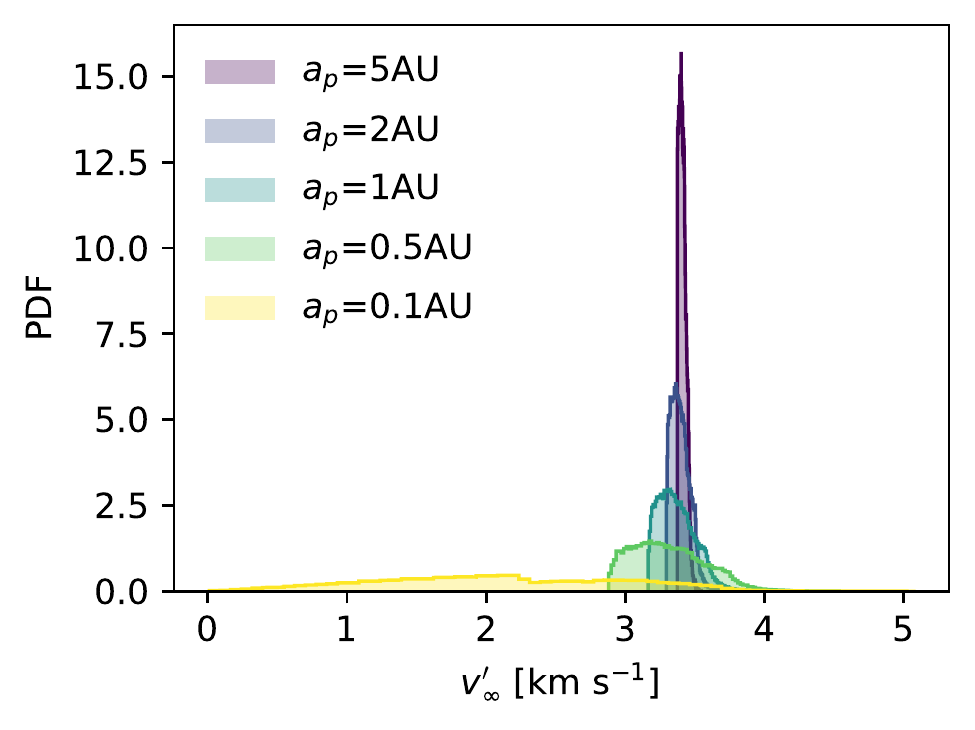}\\
	\includegraphics[width=1\columnwidth]{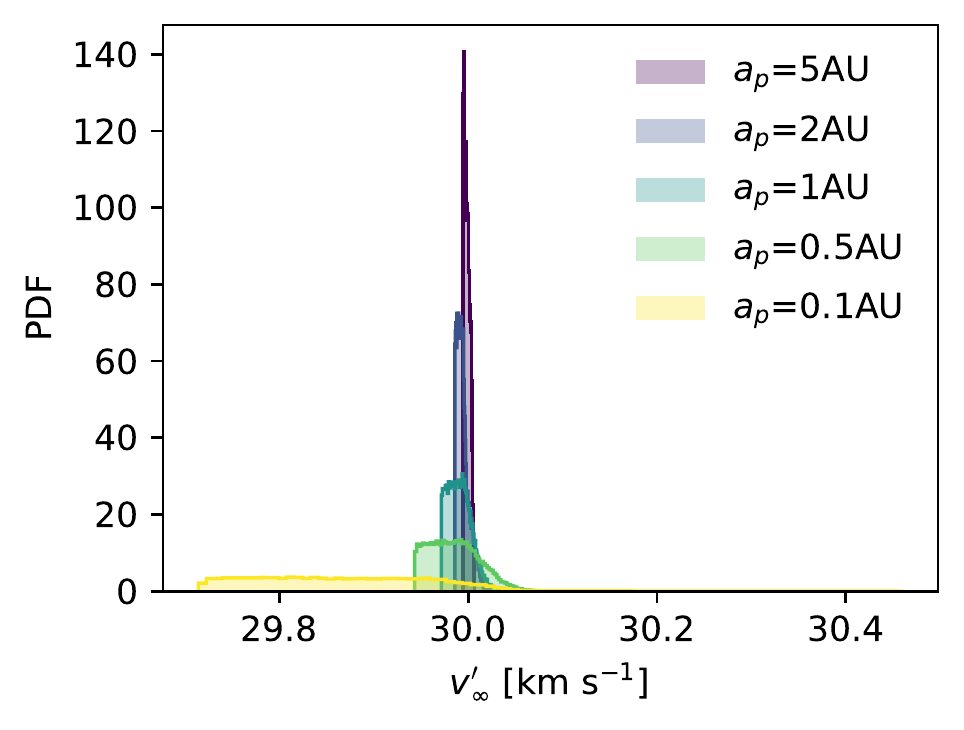}
	\caption{Post scattering relative velocity distribution, for different values of the initial approach velocity:
		\textit{Upper panel:} $v_\infty = 3.4$ km~s$^{-1}$; \textit{Bottom panel:} $v_\infty=30$ km~s$^{-1}$.}
	\label{fig:v-dist}
\end{figure}

\fig{fig:v-dist} shows the post scattering velocity of the centre of mass of the new star-planet system for different initial $v_\infty$.  Similarly as for the semi-major axis and eccentricity, both prograde and retrograde orbits contribute to the final distributions. The figure indicates that the planet transfer process can modify the velocity of the incident star by shifting the binding energy of the planet orbit from the old to the new value. The binding energy shift can be estimated as
\begin{equation}
	\Delta E = \frac{GM_{\rm p}}{2}\bigg( \frac{-M_{*2}}{\ap^\prime}-\frac{-M_{*1}}{\ap} \bigg )\,.
\end{equation}
The corresponding velocity shift then can be deduced from
\begin{eqnarray}
	&&\Delta E \sim \frac{-1}{2}\bigg(M_{*1}\Delta v_{*1}^2 + M_{*2}\Delta v_{*2}^2 \bigg)\\
	&&M_{*1}\Delta v_{*1} \sim M_{*2}\Delta v_{*2}\\
	&&\Delta v_\infty \sim  \Delta v_{*1} + \Delta v_{*2}\,.
\end{eqnarray}
Finally, we get
\begin{equation}
	\Delta v_\infty \sim  \frac{M_{*1}+M_{*2}}{M_{*1}M_{*2}}\frac{GM_{\rm p}}{2v_\infty}\bigg( \frac{M_{*2}}{\ap^\prime}  - \frac{M_{*1}}{\ap} \bigg )\,.
\end{equation}
The distribution of $\ap^\prime$ that is displayed in \fig{fig:post-properties} correspondingly gives the distribution of the new centre of mass velocity of the new planet-star system at infinity. The velocity shift can be estimated as
\begin{equation}
	\Delta v_\infty \sim \bigg( \frac{M_{*2}}{\ap^\prime}  - \frac{M_{*1}}{\ap} \bigg ) \sim \frac{AM_{*2} - M_{*1}}{\ap}\,,
\end{equation}
where $A\ap^\prime = \ap$. The '$-1$'

power law of $\ap$ explains the wider dispersion of low $\ap$ in \fig{fig:v-dist}, and the long tail of the $\ap^\prime/\ap$ distribution in the left upper panel of \fig{fig:post-properties}, where $\ap^\prime/\ap>1$, gives the long tail on the low end in \fig{fig:v-dist}.

\subsubsection{The final velocity distribution of the ejected planet}

\begin{figure*}
	\includegraphics[width=1\columnwidth]{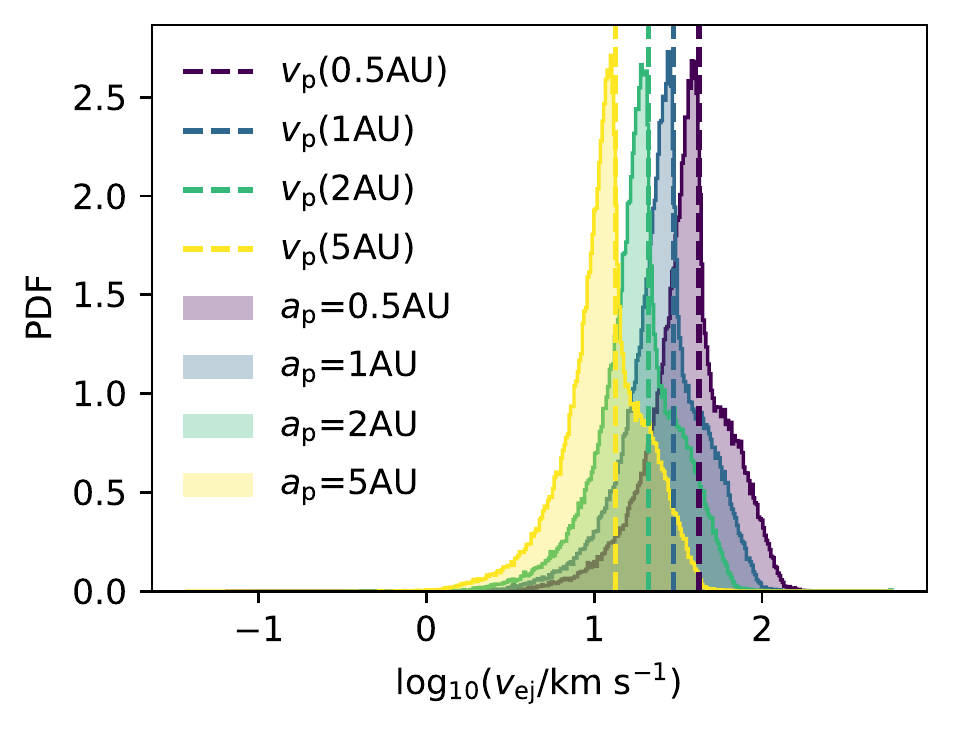}
	\includegraphics[width=1\columnwidth]{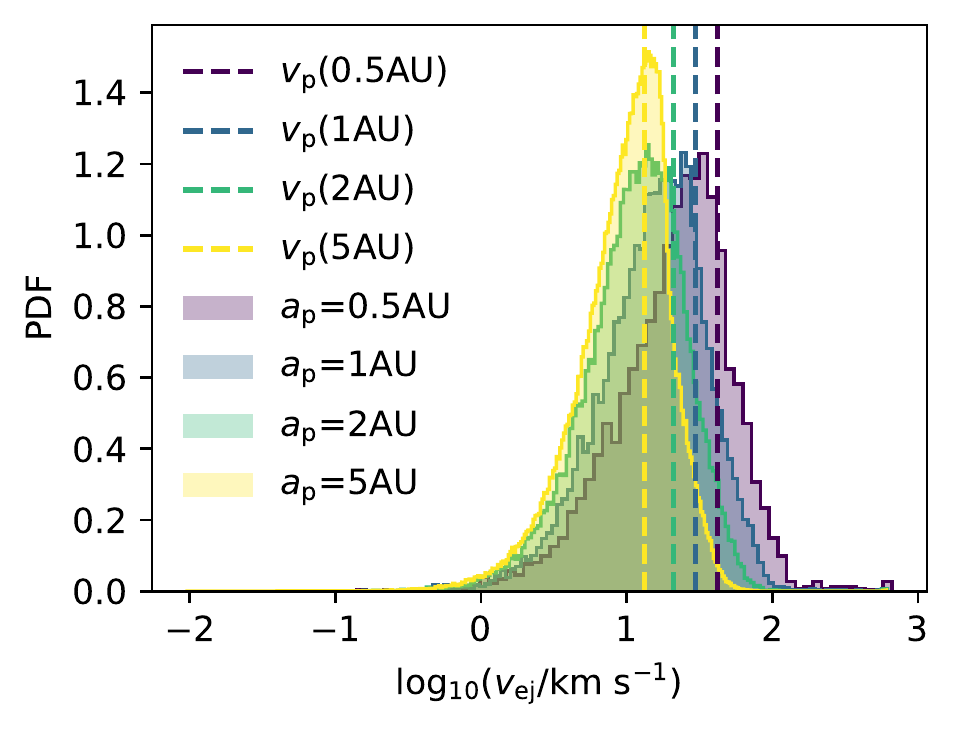}\\
	\includegraphics[width=1\columnwidth]{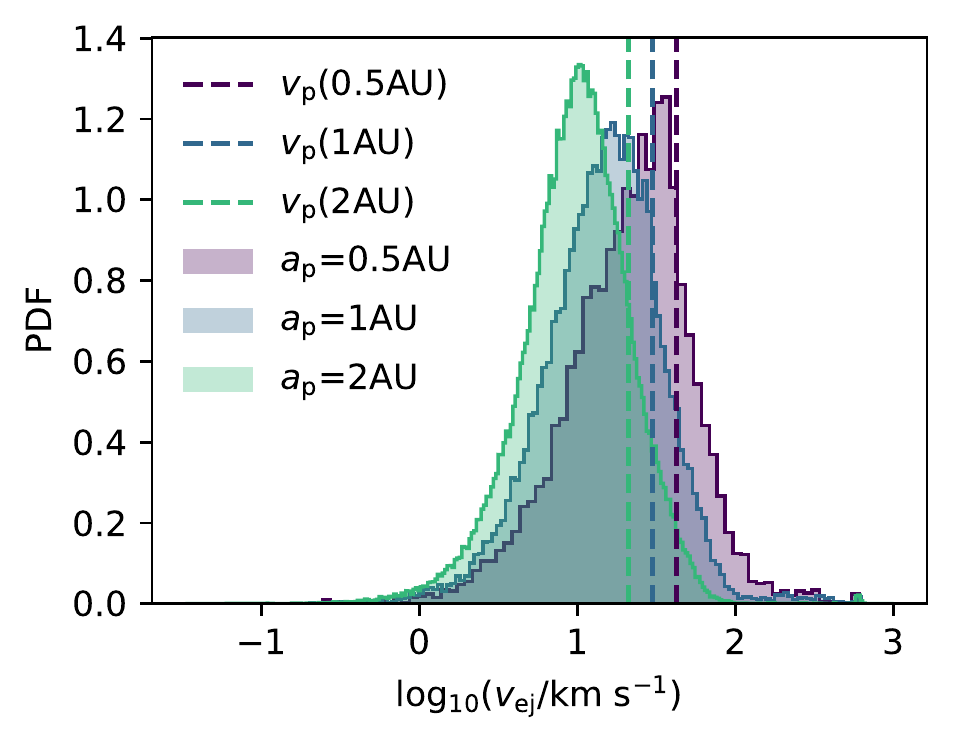}
	\includegraphics[width=1\columnwidth]{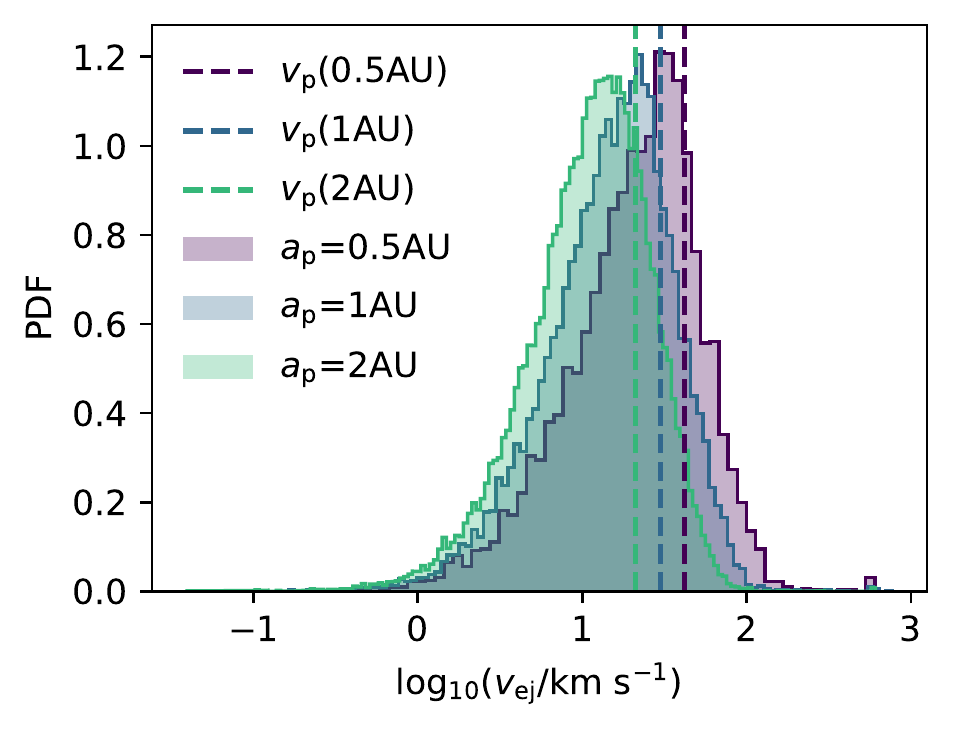}
	\caption{The velocity distribution of the ejected planet at infinity. \textit{Upper left panel:} 1+1; \textit{Upper right panel:} 2+1; \textit{Bottom left panel:} 1+2; \textit{Bottom right panel:} 2+2. }
	\label{fig:ejection-v}
\end{figure*}

The most common outcome from our scattering simulations are planet ejections. \fig{fig:ejection-v} shows the velocity distribution of the ejected planet at infinity, $v_{\rm ej}$.  The vertical lines indicate the initial orbital velocity of the planet,
\begin{equation}
	v_{\rm p}=\sqrt{\frac{GM_{*1}}{\ap}}.
\end{equation}
We find that the ejection velocity shows only a weak dependence on the velocity dispersion of the environment in which the interaction occurs (see the Appendix for more details).

The upper left panel of \fig{fig:ejection-v} shows the $v_{\rm ej}$ distribution for 1+1 scatterings.  Clearly, for different initial values of $\ap$, the $v_{\rm ej}$ distribution has similarly dual peaks around $v_0$. The right peak is due to prograde ejections.  Here, the ionization occurs such that the planet is ejected in the same direction as its original orbital velocity vector.  The left peak is due to retrograde ejections.  Here, the incident object first decelerates the planet, by accelerating it in the opposite direction.  The ionization occurs in such a way that the planet is ejected in the direction opposite to the original orbital velocity vector.

The upper right panel, bottom left panel and bottom right panel of \fig{fig:ejection-v} show the $v_{\rm ej}$ distribution for, respectively, 2+1, 1+2 and 2+2 scatterings. If a binary star is involved in the scattering, then the velocity distribution of the ejected planet has a different shape relative to the 1+1 case. This is because,
for 
scatterings 
 involving binary stars,  the disruption and recombination of the stars making up the binary (or binaries) provides an additional source of energy for planet ejection. This additional channel changes the overall shape of the velocity distribution.

 The peak of the ejection velocity distribution for the 1+2 and 2+2 cases shifts towards the left relative to the value of $v_{\rm p}$ corresponding to each value of $\ap$. Indeed, in the 1+2 and 2+2 cases, the planet's host is a star in a binary, where the host star orbits around the centre of mass of the binary. This additional motion changes the relative velocity between the planet and the intruder, hence leading to changes in the ejection velocity distribution. Due to the isothermally distributed angles in both the planet and the binary star orbits, the relative velocity between the planet and the intruder can in principle both increase and decrease. However, the numerical experiments indicate that this additional motion produces an overall shift of the ejection velocity distribution towards lower velocities and  widens the dispersion.

Notice that in the 1+2 and 2+2 scatterings, there is no $\ap=5$~AU case. The is because the binary host in the 1+2 and 2+2 cases has a separation of 10 AU. A planetary orbit with $\ap=5$~AU becomes unstable in such a system.

\subsection{Outcome cross sections and rates in different environments}

In the following, we use our results to calculate the cross section and the event rate of planet transfers, planet ejections and collisions for specific astrophysical environments, i.e, open clusters, globular clusters and the galactic field.
These environments differ in their stellar densities, and hence in the corresponding velocity distributions of the stars.
For virialized open clusters and globular clusters, the velocity dispersion and number density are correlated. We adopt the virialized model
\begin{equation}
n_{\rm vir}\sim\frac{M_{\rm c}/\bar{m}}{4\pi R_{\rm c}^3/3}\sim\frac{6v^6}{\pi G^3M_{\rm c}^2\bar{m}}\,,
\end{equation}
where $v$ is the mean star velocity of the cluster, $M_{\rm c}$ its total mass and $\bar{m}$  the mean stellar mass.
For the open cluster case, we adopt a velocity dispersion $\sigma=1$~km~s$^{-1}$ with typical cluster mass $M_{\rm c}\sim 2\times10^3$~M$_{\odot}$ \citep{Kroupa02}; for globular clusters we adopt the velocity dispersion $\sigma=5$~km~s$^{-1}$, with typical cluster mass $M_{\rm c}\sim 4.5\times 10^4$~M$_{\odot}$ \citep{Kimmig15}. Using a typical cluster mean stellar mass $\bar{m}\sim$~0.5~M$_{\odot}$, we get $n_{\rm vir}\sim$3.5$\times 10^3$~pc$^{-3}$ and  $\sim$1.1$\times 10^5$~pc$^{-3}$ for open and globular clusters, respectively. For  field stars, the correlation between $n_{\rm vir}$ and $v$ cannot be applied. For this case we adopt the typical values $\sigma=30$~km~s$^{-1}$ and $n\sim1$~pc$^{-3}$. The planet separation in open clusters and the galactic field is set to be 1~AU, while it is 0.1~AU in globular clusters \citep{Cai2019}. The binary star separation is set to be 10~AU.

\begin{figure*}
	\includegraphics[width=2\columnwidth]{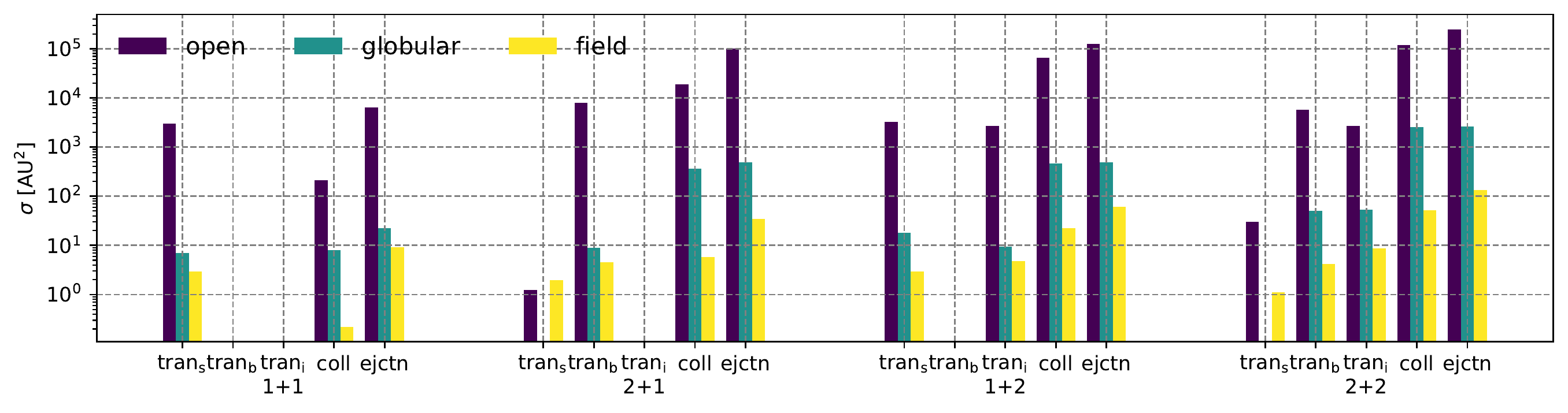}
	\includegraphics[width=2\columnwidth]{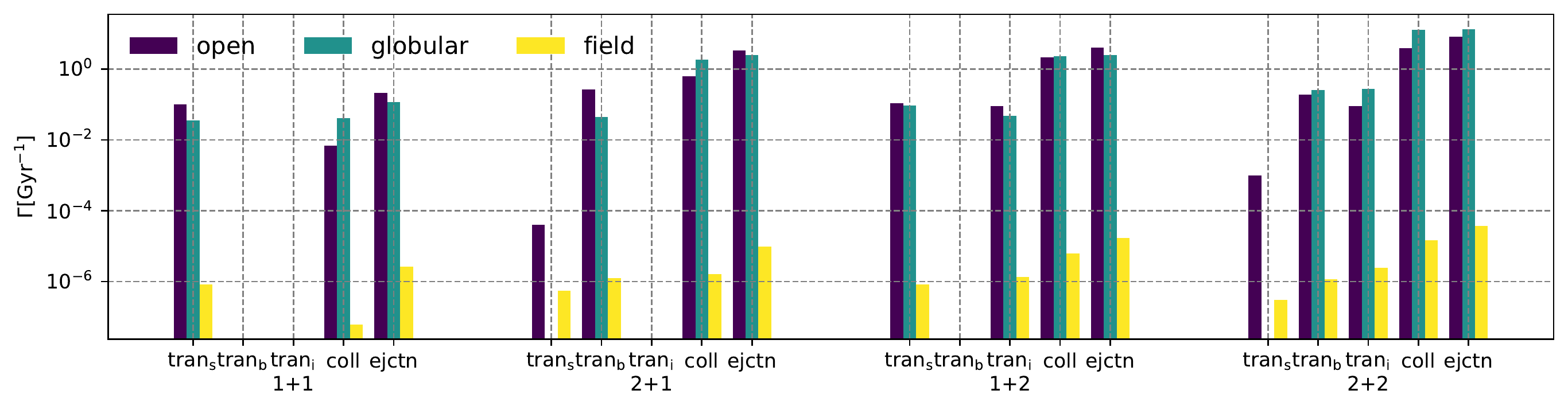}
	\caption{\textit{Upper panel:} Cross section of outcomes in open clusters, globular clusters and the galactic field. \textit{Bottom panel:} Interaction rate of outcomes in open clusters, globular clusters and the galactic field.}
	\label{fig:evn}
\end{figure*}

The upper panel of \fig{fig:evn} shows the cross sections for planet transfers, collisions (star-star and star-planet) and planet ejections for 1+1, 1+2, 2+1 and 2+2 scatterings under different environmental conditions. The bottom panel shows the corresponding rates 
\begin{equation}
	\Gamma_{\rm x} = \sigma_{\rm x} n  v \,,
\end{equation}
where $\sigmax$ is the cross section of outcome $x$, $n$ is the number density and $v$ is the mean velocity of the environment given by
\begin{equation}
	\sigma=\sqrt{\frac{3\pi -8}{3\pi}}v\,,
\end{equation}
with  $\sigma$ being the velocity dispersion. The bottom panel shows how, across all the different environments, the cross section for planet ejection is the largest, implying that free-floating planets should be quite common in interacting stellar environments. Collisions are very dominant in the 2+1, 1+2 and 2+2 cases as well, especially in globular clusters. Notice that for the 1+1 case in open  and globular clusters, the rate of planet transfers becomes comparable to that of collisions and ejections; thus a planet in these environments has a very large probability to not orbit around its mother star. For the 1+2 and 2+2 cases (where the planet originally orbits one of the components of a binary), the transfer probability is much smaller due to the relatively larger ejection rates, which indicates that we have a larger probability to observe a planet orbiting around a single star rather than one component of a binary star, if stellar interactions are indeed frequent in these environments. For galactic field stars, the low rate indicates that the stellar interactions are rare; this is due to the large velocity dispersion of the stars. In these galactic field environments, peculiar planetary architectures resulting from dynamical interactions could be the remnants of earlier phases of interactions of the star if it formed within a group.



\section{Summary and Conclusions}

In this work we have studied the effects of dynamical interactions to planetary architectures in interacting stellar environments. 
The systems we considered are composed of a giant planet orbiting either a single star or one of the components of a binary star system, and a close flyby by either a single star or a binary.

By means of high-precision N-body simulations
we have estimated the cross sections and rates of planetary ejections, transfers and collisions for single-single stars, single-binary stars and binary-binary  interactions. Our main conclusions can be summarized as follows:

\begin{itemize}
\item{}
Interactions between binary star-single star and binary-binary can significantly increase (by one order of magnitude) the cross section for planet ejection compared with single-single star interactions. In interacting stellar environments, although the binary fractions tend to be lower, there will however be relatively large amounts of free-floating planets.

\item{}
The cross section for planet transfers (from one star to another star) is non-negligible in interacting stellar environments. The
distribution of 
semi-major axes of the transferred planet is broad, with a peak which depends on the mass ratio between the original and the new star; the new orbits display a high eccentricity.
Especially for single-single star interactions, the cross section for planet transfers is comparable to the cross section for ejections. Therefore, in the interacting environments which typically have a  low binary fraction, planet transfers can be relatively common.

\item{}
Unlike planet ejections and  collisions for which binary interlopers can significantly increase the cross section, 
the cross section for planet transfer shows a slight dependence on whether the flyby is by a single star or by a binary.
Therefore, for an environment with different binary fractions, the averaged cross section for planet transfer is similar. The cross section for planet transfer shows only a dependence on the star masses and their velocity dispersion.

\item{}
The ejection velocity of the planet is indicative of the original orbital velocity of the planet, and thus of the original semi-major axis of the planet. In single-single star interactions, planets with different semi-major axes show identical ejection velocity distributions around the orbital velocity. In  interactions involving binaries, slower ejection velocities (dependent on the binary star semi-major axis) will be produced. 
\end{itemize}

\section*{Acknowledgements}

NWCL gratefully acknowledges the support of a Fondecyt Iniciaci\'on grant \#11180005.


\bibliography{refs}



\appendix
\section{Additional Figures}

Here we display further results of our simulations, which extend the parameter space explored in the main paper. 

\begin{figure*}
	\includegraphics[width=\columnwidth]{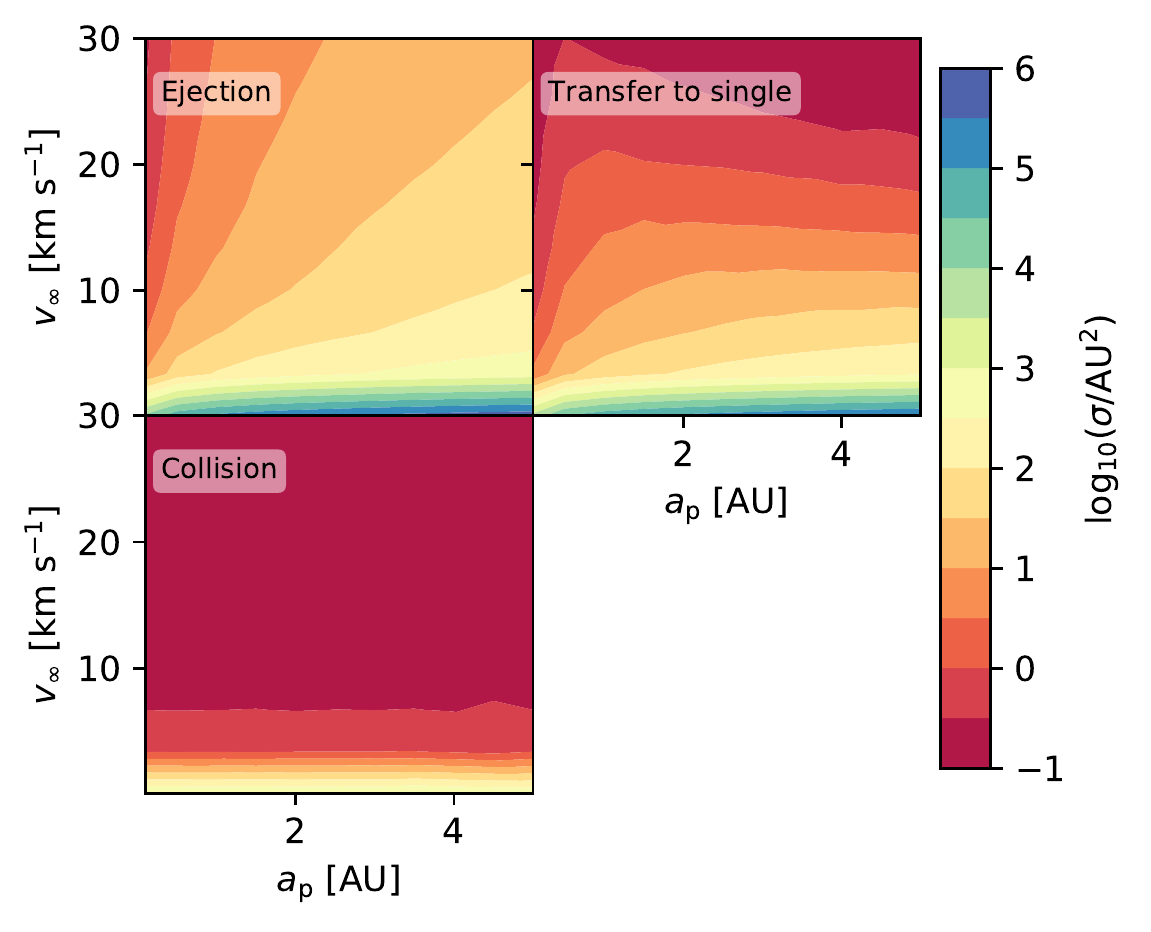}
	\includegraphics[width=\columnwidth]{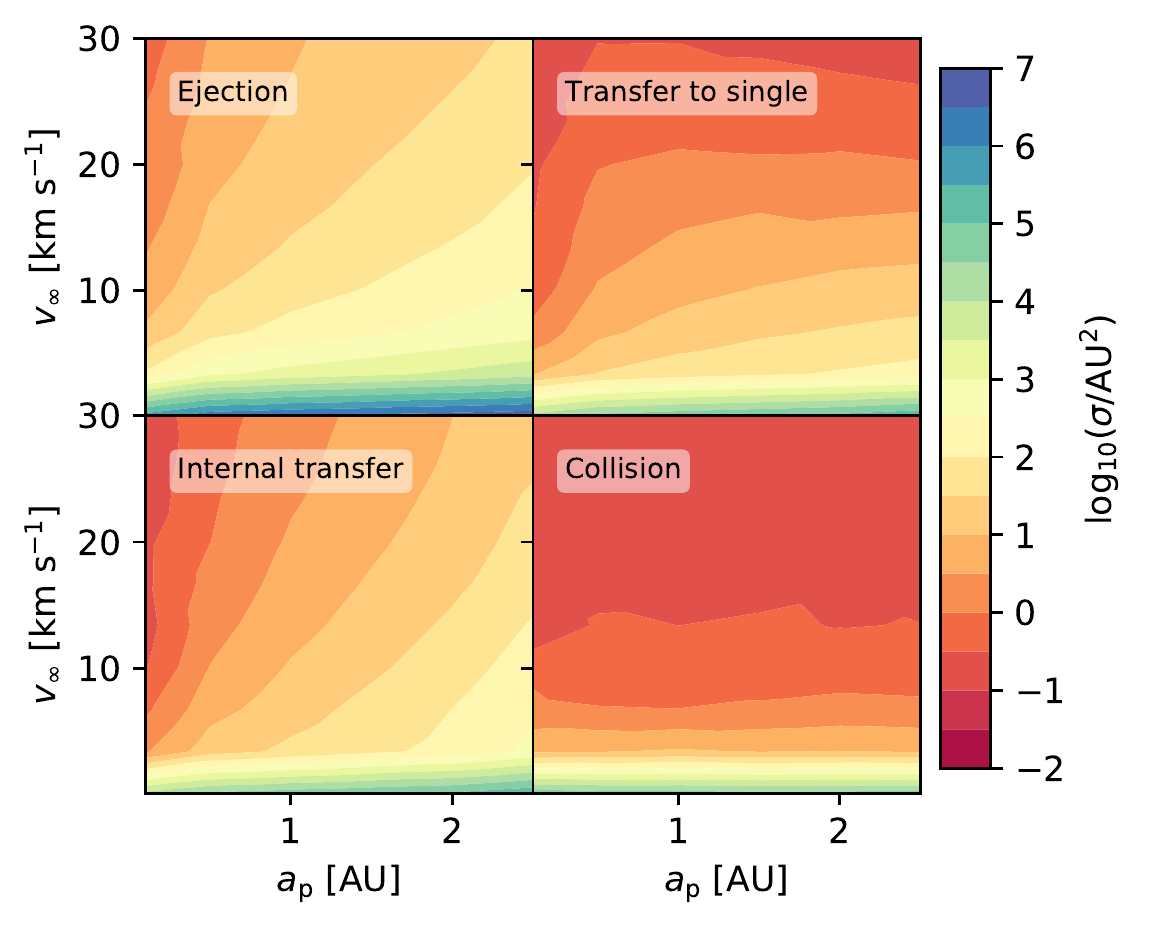}\\
	\includegraphics[width=\columnwidth]{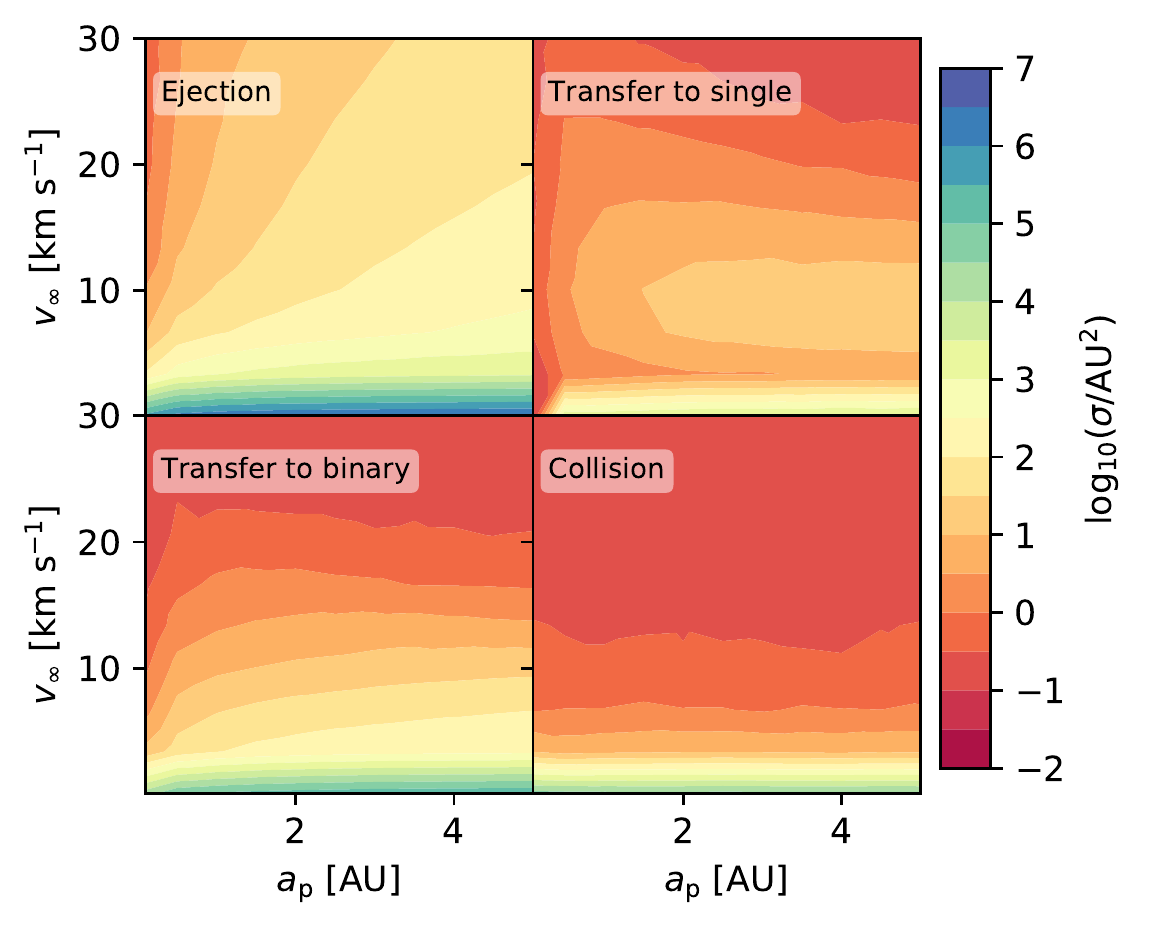}
	\includegraphics[width=\columnwidth]{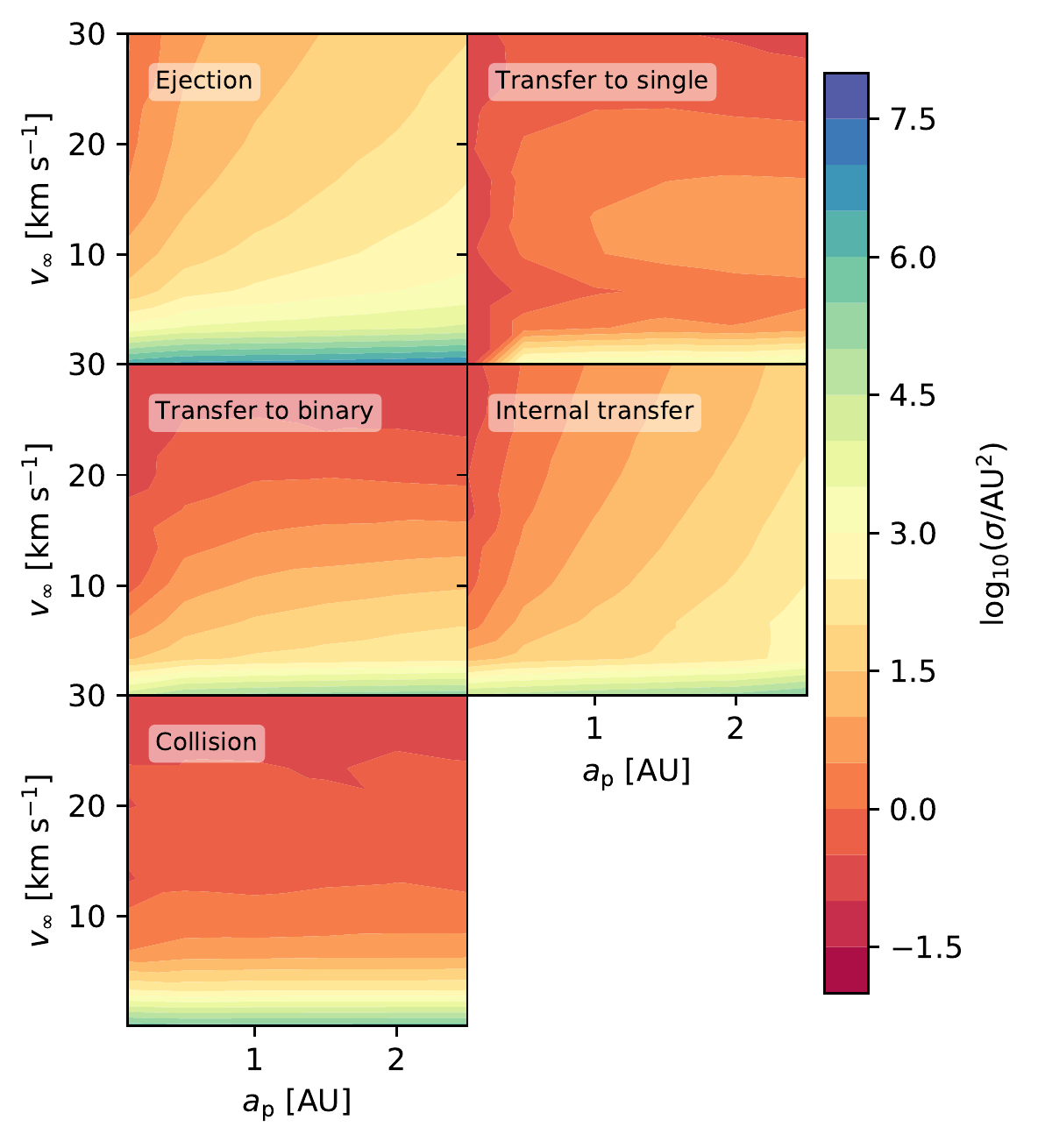}
	\caption{Cross section for planet ejections, planet transfers, and collisions for 1+1 (upper left), 1+2 (upper right), 2+1 (bottom left) and 2+2 (bottom right) scatterings as a function of $v_\infty$ and $\ap$. The mass of the star originally hosting the planet is $M_{*1}=1\,M_\odot$, while the mass of the interloper is $M_{*2}=0.2M_\odot$ for a single star and for each member star, if a binary; the mass of the planet is $M_{\rm p}$=1$M_{\rm J}$ and the binary star separation is $\as$=10~AU.}
	\label{fig:A1}
\end{figure*}

In particular, we study the dependence on the interloper mass in Fig.~\ref{fig:A1}, by running experiments with the star mass equal to $M_{*,2}=0.2~M_\odot$ for the case of a single star, and equal to $M_{*,2}=0.2~M_\odot$ for each component in case of a binary flyby.
Fig.~\ref{fig:A1}  shows a cross section about an
order of magnitude 
lower in comparison to \fig{fig:1+1}, \fig{fig:2+1}, \fig{fig:1+2} and \fig{fig:2+2}, due to the lower intruder mass. Furthermore, we note that in the 2+1 and 2+2 cases the critical velocity of transfer to single ($v_{\rm orb, i}=\sqrt{2GM_{*2}/a_{\rm*}}$) drops from $\sim$ 10~km~s$^{-1}$ to $\sim$ 5~km~s$^{-1}$. Indeed, this critical velocity  divides the parameter space into two regimes where the binary intruder tends/ does-not-tend to be disrupted. To be transferred to an isolated single star, disruption of the binary intruder is required. Thus, we can see the difference in \fig{fig:A1} for low $M_{*,2}$ due to the shifts in $v_{\rm orb, i}$.  For other cases, we do not see a significant difference in the shape of the contour plots.

\begin{figure*}
	\includegraphics[width=2\columnwidth]{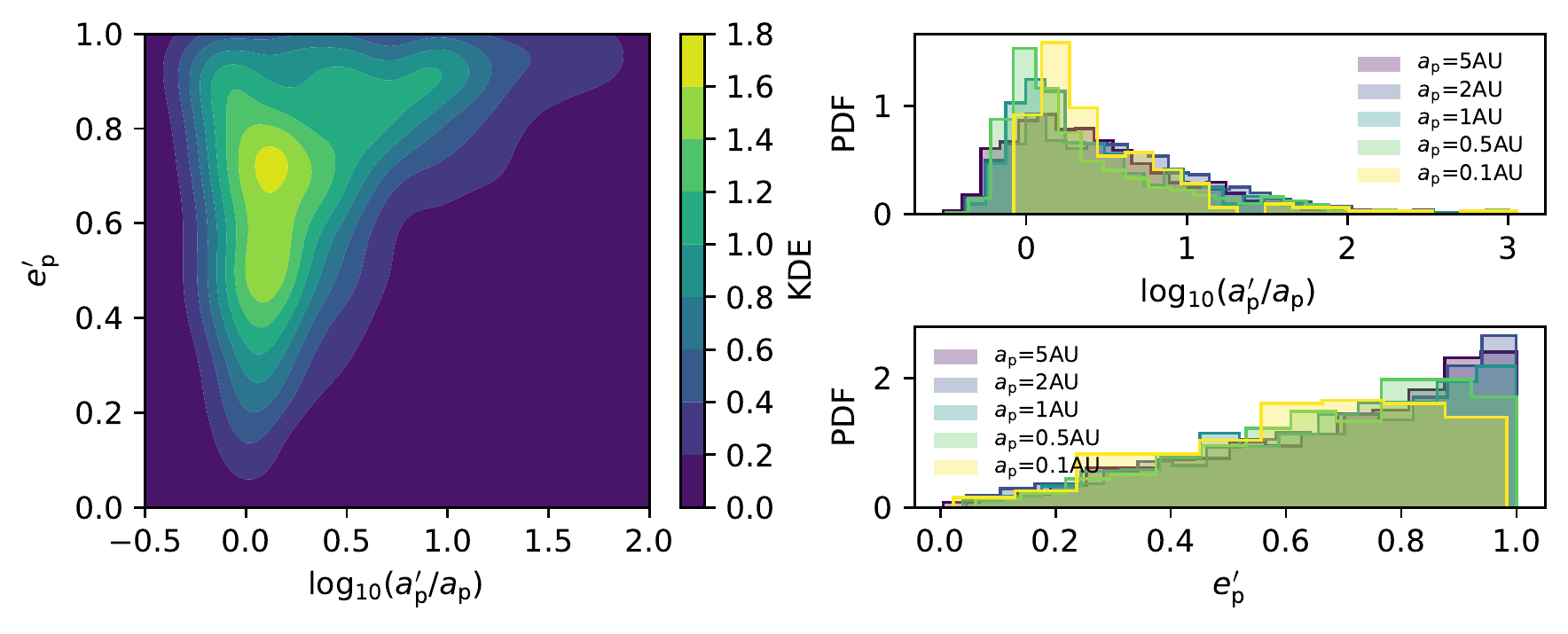}
	\includegraphics[width=2\columnwidth]{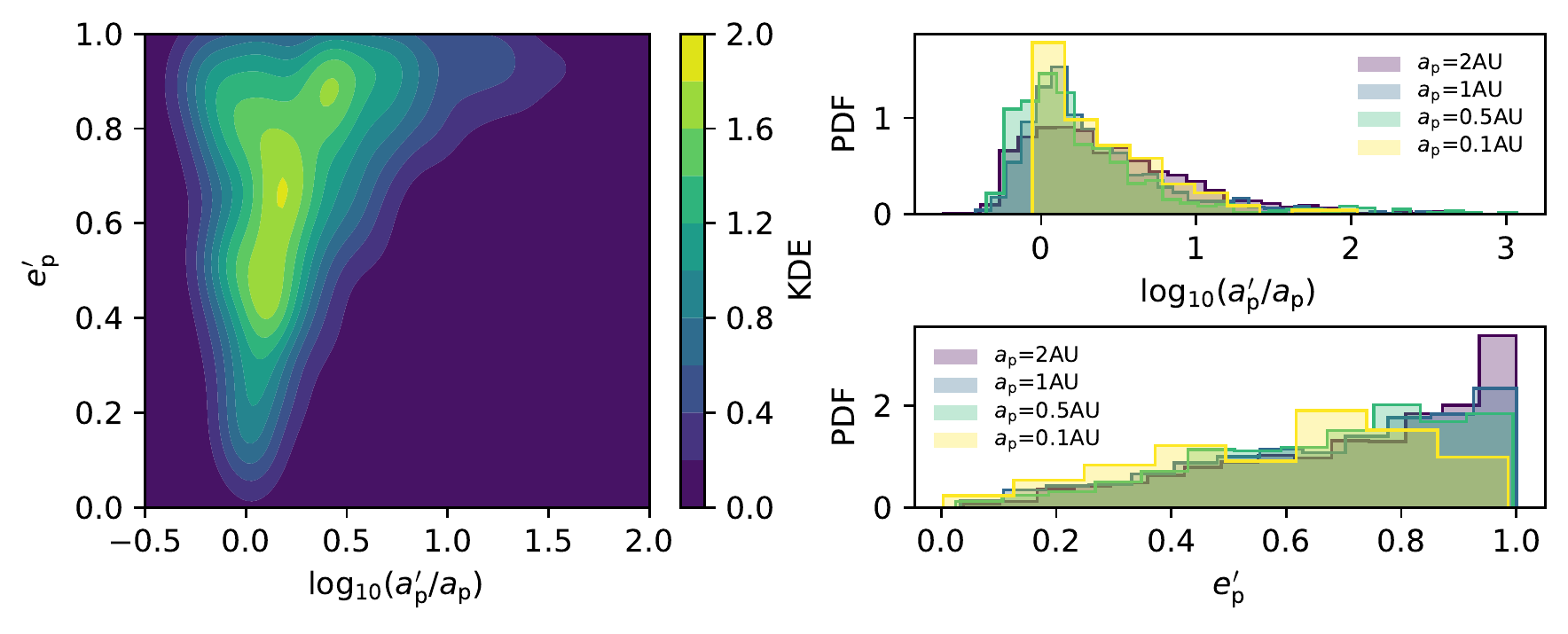}
	\includegraphics[width=2\columnwidth]{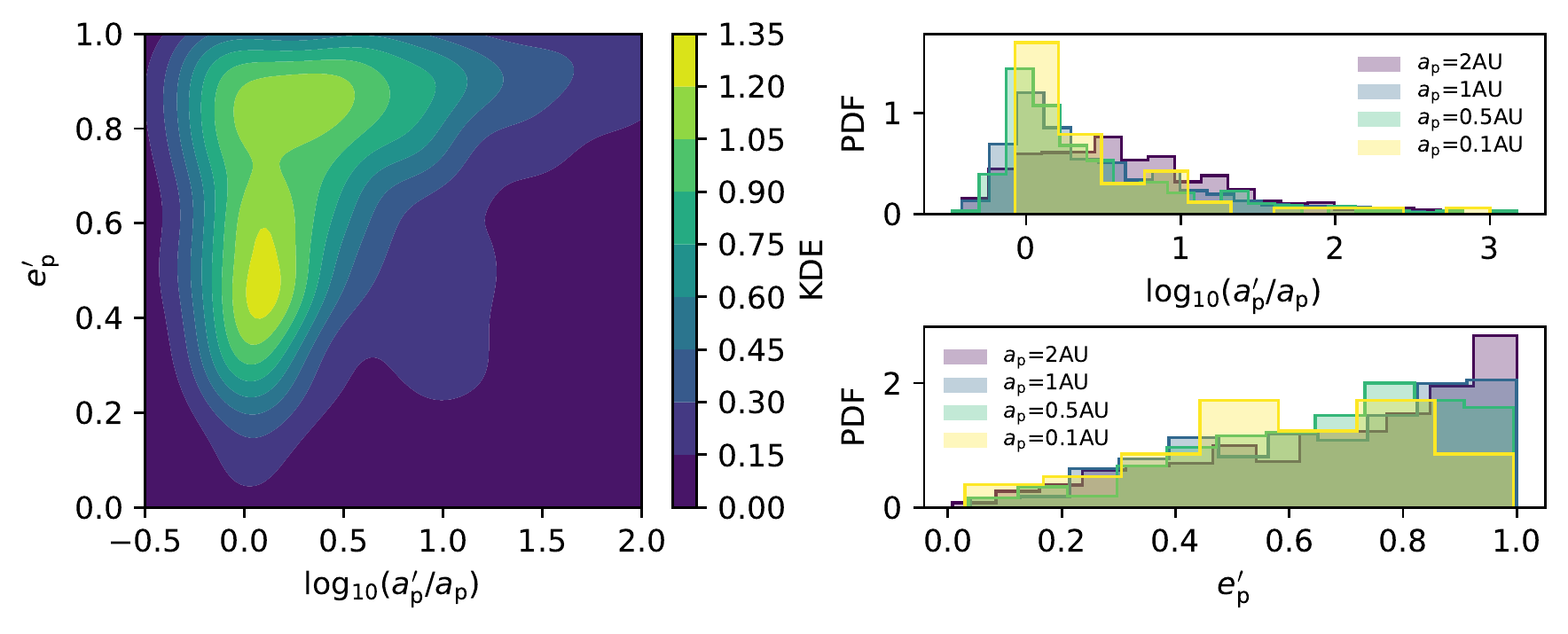}
	\caption{Orbital properties of the externally transferred planet for the 2+1, 1+2 and 2+2 cases. The final properties show a weak dependence on the interaction type. }
	\label{fig:A2}
\end{figure*}

Fig.~\ref{fig:A2} extends the study of the orbital properties of the externally transferred planet (limited to the 1+1 case in the main body of the paper), to the cases in which the planet is initially orbiting one member of a binary (1+2 and 2+2), or a single star (2+1) and is transferred onto either a single star (1+2) or to one of the members of an interloper binary (2+1 and 
2+2 cases). 

\begin{figure*}
	\includegraphics[width=2\columnwidth]{figs/post-properties01}
	\includegraphics[width=2\columnwidth]{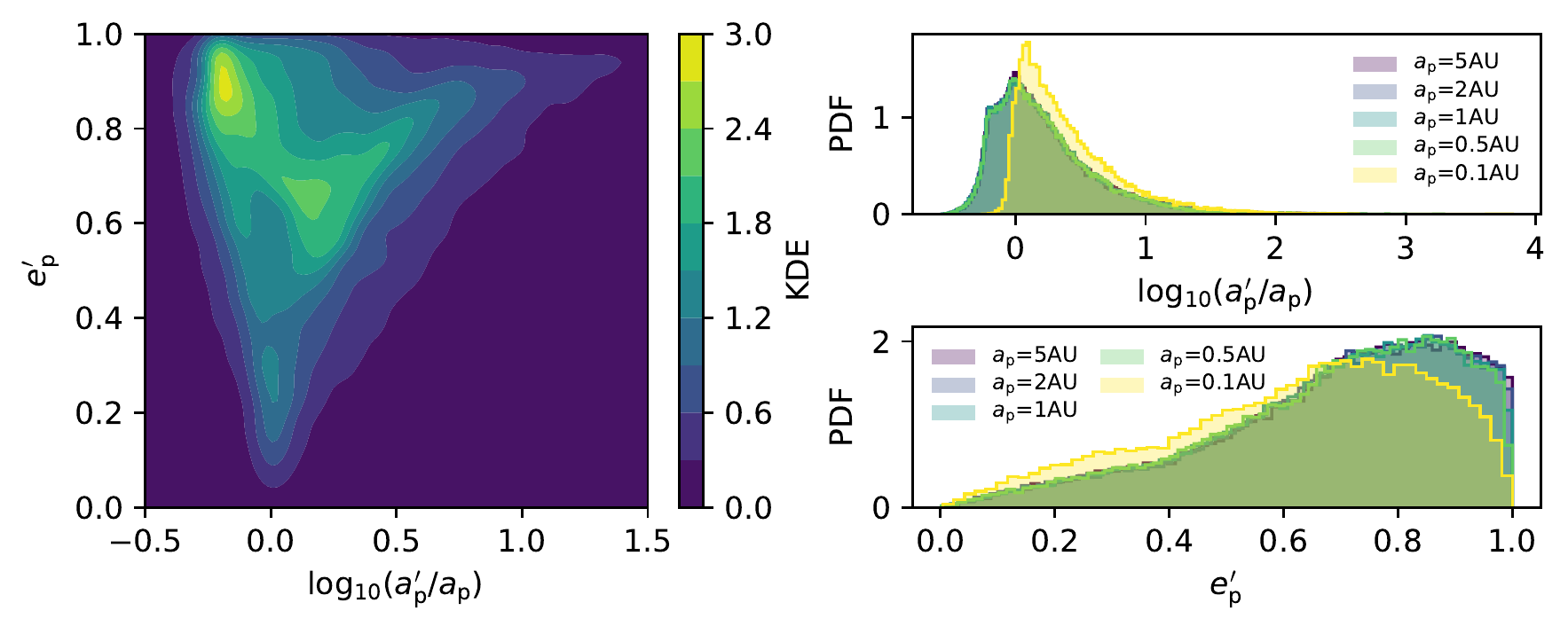}
	\includegraphics[width=2\columnwidth]{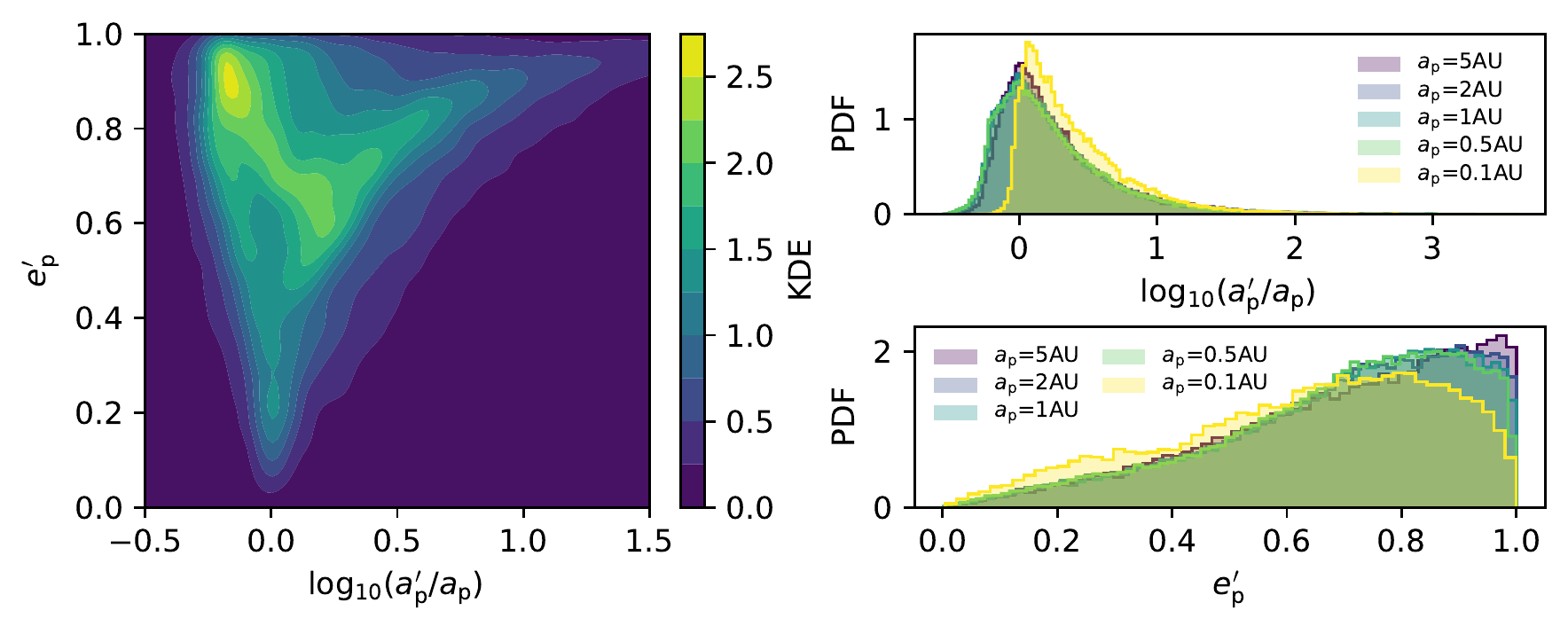}
	\caption{Orbital properties of the external transferred planet for the 1+1 case and $v_\infty=0.1$~km~s$^{-1}$(upper panel), $3.4$~km~s$^{-1}$ (middle panel) and $10.1$~km~s$^{-1}$ (bottom panel). The final properties show a weak dependence on $v_\infty$.}
	\label{fig:A3}
\end{figure*}

\begin{figure*}
	\includegraphics[width=2\columnwidth]{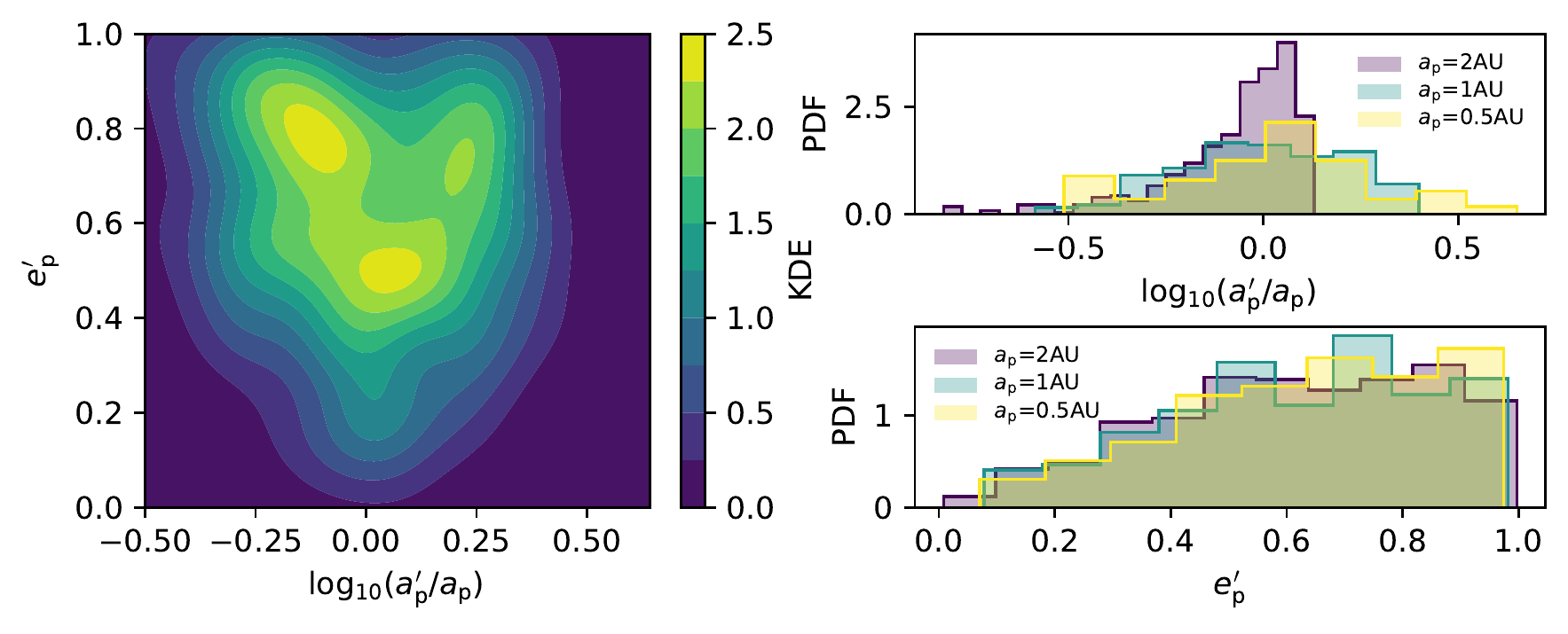}
	\includegraphics[width=2\columnwidth]{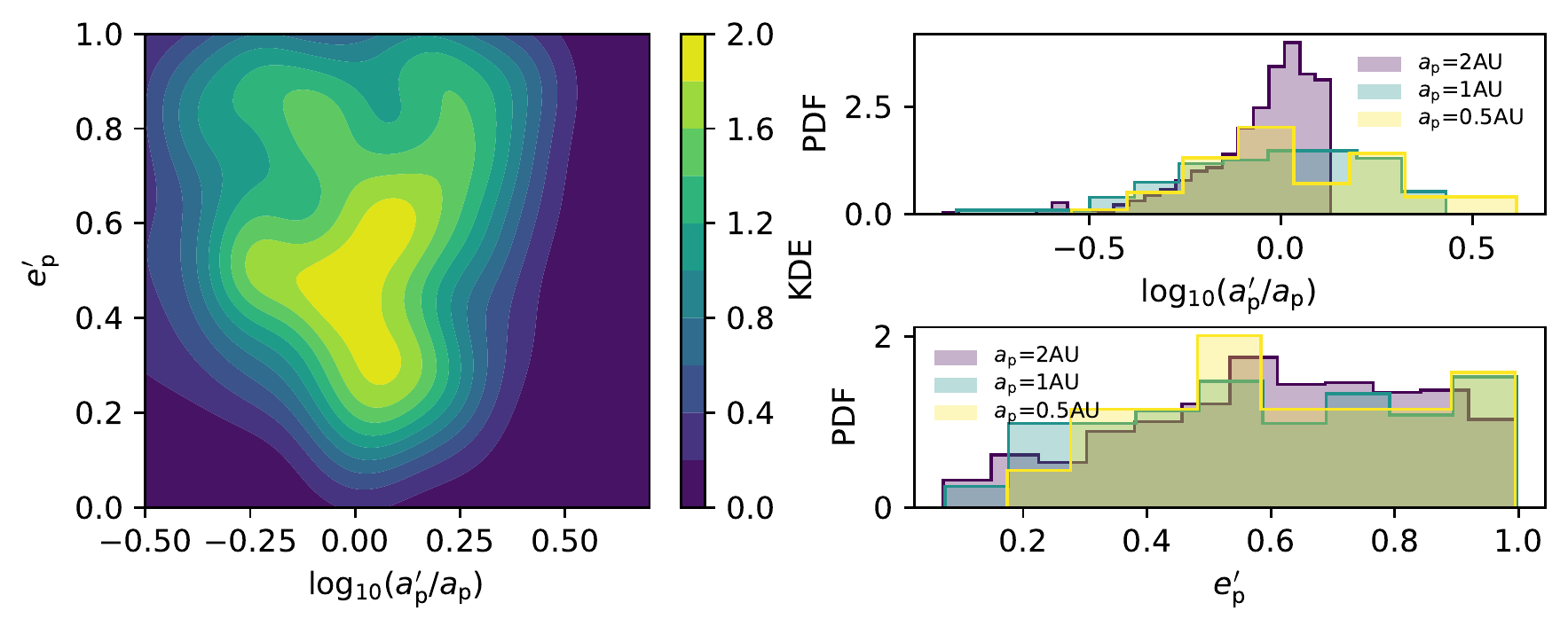}
	\includegraphics[width=2\columnwidth]{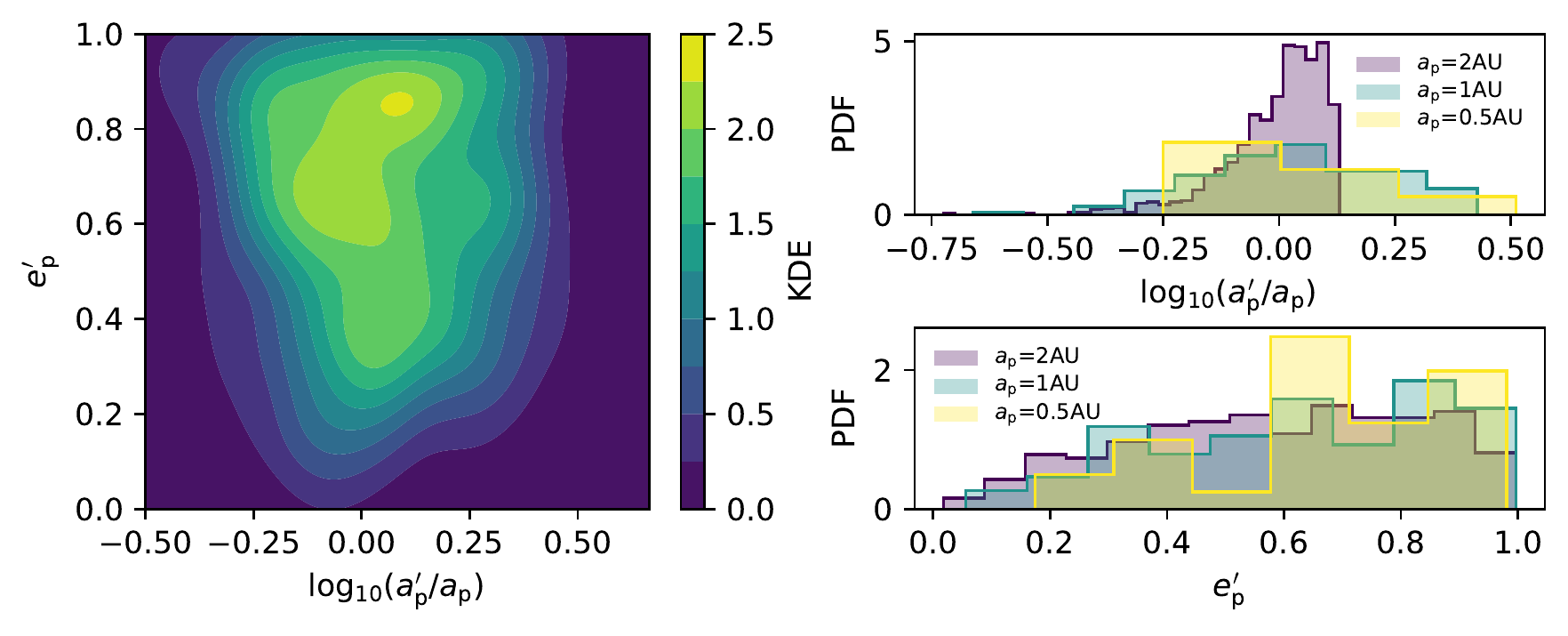}
	\caption{Orbital properties of the internal transferred planet for the 1+2 case for $v_\infty=0.1$~km~s$^{-1}$ (upper panel), $3.4$~km~s$^{-1}$ (middle panel) and $10.1$~km~s$^{-1}$ (bottom panel). The final properties show a weak dependency on $v_\infty$.}
	\label{fig:A4}
\end{figure*}

\begin{figure*}
	\includegraphics[width=2\columnwidth]{figs/itrans-bs-post-properties34}
	\includegraphics[width=2\columnwidth]{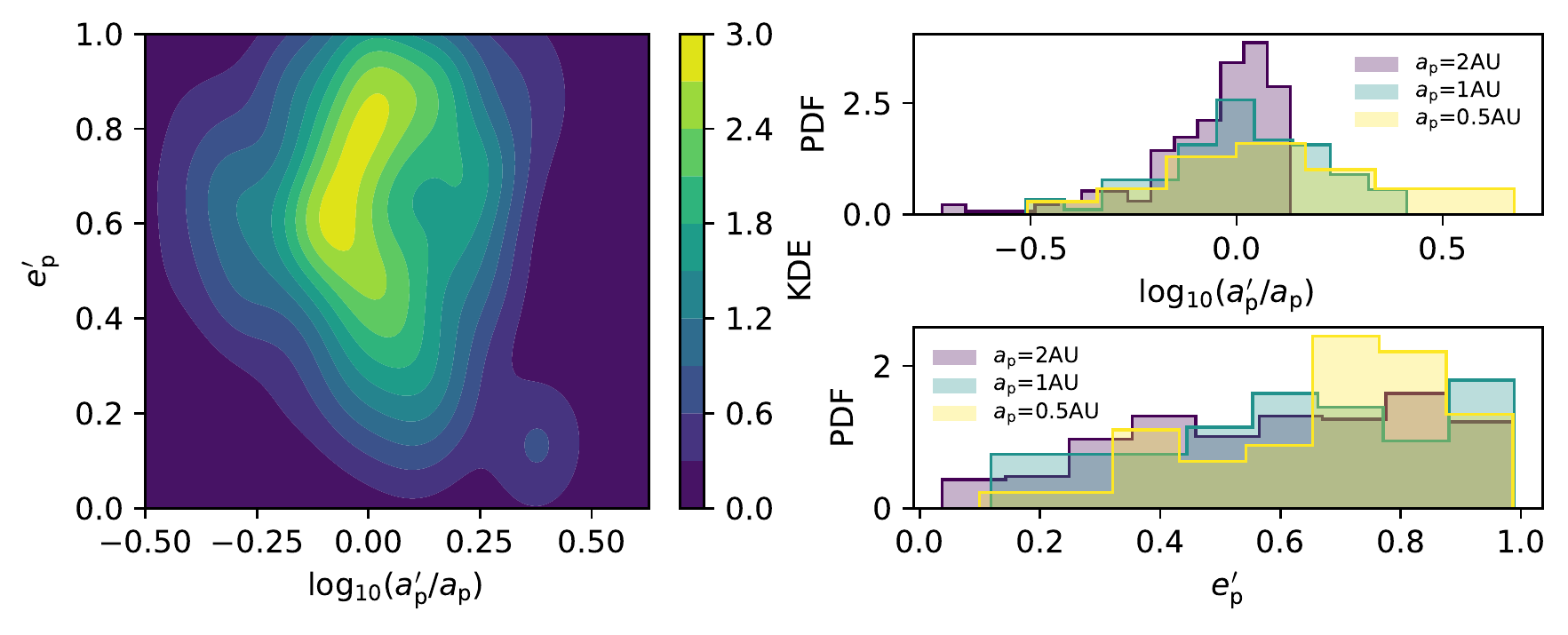}
	\caption{Orbital properties of the internal transferred planet for the 1+2 (upper panel) and 2+2 (bottom panel) cases with $v_\infty=3.4$~km~s$^{-1}$. The final properties show a weak dependence on the  interaction type.}
	\label{fig:A5}
\end{figure*}

The dependence of the orbital properties of the planet for the external transfer 
in the 1+1 case are displayed in Fig.~\ref{fig:A3} for three different values of the relative star velocity at infinity.  Similarly, Fig.~\ref{fig:A4}
shows the velocity dependence of the orbital parameters of the exchanged planet for the internal (1+2) case. 
Last, Fig.~\ref{fig:A5} shows the same orbital distributions of the transferred planet for a fixed value of the incoming velocity, but for two different interaction types (1+2 and 2+2). 
As can be seen by comparing the various cases, the dependence of the distributions on both the relative scattering velocities and the interaction type is rather weak. 


\bsp	
\label{lastpage}
\end{document}